\documentclass[useAMS]{mn2e}
\usepackage{mnras_cite}
\usepackage{epsfig}
\usepackage{color}
\usepackage[normalem]{ulem}
\usepackage{graphicx}
\usepackage{subfigure}
\usepackage{multirow}

\newcommand{\dif}{{\rm d}}
\newcommand{\eq}{\begin{equation}}
\newcommand{\qe}{\end{equation}}
\newcommand{\ar}{\begin{eqnarray}}
\newcommand{\ra}{\end{eqnarray}}
\newcommand{\Msun}{{\rm M}_\odot}
\newcommand{\kms}{\,{\rm km}\,{\rm s}^{-1}}
\def\ltsim{\lower.5ex\hbox{$\; \buildrel < \over \sim \;$}}
\def\gtsim{\lower.5ex\hbox{$\; \buildrel > \over \sim \;$}}

\begin{document}

\title[Principles of supernova-driven winds]{Principles of supernova-driven winds} 
\author[M.~J.~Stringer et al. ]{M.~J.~Stringer$^{1,2,3}$, R.~G.~Bower$^1$, S.~Cole$^1$, C.~S.~Frenk$^1$ and T.~Theuns$^{1,4}$\\
$^1$Institute for Computational Cosmology, Department of Physics, University of Durham, South Road, Durham DH1 3LE\\
$^2$Observatoire de Paris, LERMA, 61 Av. de l'Observatoire, 75014 Paris, France\\
$^3$Kavli Institute for Cosmology, Madingley Road, Cambridge, CB3 0HA\\
$^4$Department of Physics, University of Antwerp, Campus Groenenborger, Groenenborgerlaan 171, B-2020 Antwerp, Belgium
}

\maketitle
\begin{abstract}
The formation of galaxies is regulated by a balance between the supply of gas and the rate at which it is ejected. Traditional explanations of gas ejection equate the energy required to escape the galaxy or host halo to an estimate for the energy yield from supernovae. This yield is usually assumed to be a constant fraction of the total available from the supernova, or is derived from the assumption of a consistent momentum yield.  By applying these ideas in the context of a cold dark matter cosmogony, we derive a 1st-order analytic connection between these working assumptions and the expected relationship between baryon content and galaxy circular velocity, and find that these quick predictions straddle recent observational estimates. To examine the premises behind these theories in more detail, we then explore their applicability to a set of gasdynamical simulations of idealised galaxies.  We show that different premises dominate to differing degrees in the simulated outflow, depending on the mass of the system and the resolution with which it is simulated. Using this study to anticipate the emergent behaviour at arbitrarily high resolution, we motivate more comprehensive analytic model which allows for the range of velocities with which the gas may exit the system, and incorporates both momentum and energy-based constraints on the outflow. Using a trial exit velocity distribution, this is shown to be compatible with the observed baryon fractions in intermediate-mass systems, but implies that current estimates for low-mass systems can not be solely accounted for by supernova winds under commonly-held assumptions.
\end{abstract}

\section{Introduction}\label{Introduction}

Any viable theory of the formation and evolution of galaxies should be able to account for the mass of baryons contained, or rather not contained, in the massive collapsed regions that host galaxies. Observational constraints on the location of baryons in the Universe imply that the fraction  within these `halos' can be many times less than the cosmic baryon fraction, $f_b\approx 0.17$ (e.g. \pcite{WMAP}), and that the extent of the deficit is clearly dependent on the host's mass. This can be seen from the estimated baryonic and total masses from seven separate surveys which were collected together in one figure in the review by \scite{McGaugh10}; data which are reproduced here in our Fig. \ref{McGaugh}.

The established explanation for this defecit, dating from long before such observational data were available, is that baryons can be driven from the galaxies - and their host halos -  by supernovae explosions \cite{Matthews71}. This account is based on the premise that the energy required to escape the galaxies' gravity is readily available from the supernovae. Because the gravitational potential barrier will increase with host halo mass, the fraction of the supernova-driven wind which escapes might intuitively be expected to be greater for lower mass systems, and this does indeed seem to be qualitatively upheld by the mass-dependance seen in the modern data. 

A more quantitative version of this theory was then developed by \scite{Larson74}, who equated this potential barrier with an estimate of the energy yield per supernova (and hence per mass of stars formed). In \S\ref{Background} we review the arguments in this classic theory and, by updating the basic premises to include a cold dark matter component in the halos, show how it leads to the 1st order theoretical predictions for baryon fractions which are overlaid with the observational estimates in Fig. \ref{McGaugh}. We also take the opportunity to contrast the scaling expected from the traditional assumption, of consistent energy conversion to the ejected material (\S\ref{EnergyConversion}), with the alternative working assumption of a consistent {\em momentum} yield (\S\ref{MomentumConservation}). 

We then go on, in \S\ref{Results}, to investigate how modern {\em simulations} of disk galaxies relate to these analytic theories, using aspects of the theory to understand the behavior which emerges from these numerical experiments, depending on the parameters of the simulated supernovae and the resolution with which the system is simulated. This investigation is then used to motivate revisions to the traditional premises and in \S\ref{Picture} we present a alternative derivation of the outflow which is built on more realistic assumptions, whilst keeping to the spirit of the original elegant theories.

\section{Theoretical Background}\label{Background}
\subsection{Consistent energy conversion}\label{EnergyConversion}

The earliest attempt to quantify the mass outflow due to supernova was by \scite{Larson74}. This brought the idea of \scite{Matthews71} together with work on the evolution of supernova remnants \cite{Cox72} to estimate that 10\% of the original supernova energy will be retained as thermal and kinetic energy in the gas, a figure calculated by considering the energy that would still be contained within supernova remnants at the point at which they begin to overlap. 

This point naturally depends on the conditions in the interstellar medium, but by including two key properties (star formation rate and gas density) as variables in the derivation, the final estimate was shown to be sufficiently insensitive to variations in these conditions that it was considered, ``under most circumstances likely to be encountered in practice'', accurate ``to within about a factor of 2''.

This figure could then be incorporated into their model of gas ejection under the additional premise:
\begin{quotation}
``...that all of the available energy of $~0.1E_{\rm SN}$ is expended in removing gas from the galaxy." \cite{Larson74} 
\end{quotation}
This premise can be expressed as the assumption of {\bf consistent energy conversion},  in that the energy yield is consistent from one system to another and is {\em all} converted to the gravitational potential energy gained by the gas in escaping the halo. This amounts to the relation:
\eq
M_{\rm esc}v_c^2 \approx \eta E_{\rm SN}~, \label{Larson}
\qe
where $M_{\rm esc}$ is the mass of ejected gas, $v_{\rm c}$ is the characteristic velocity of the system (defined, for example, as the maximum circular velocity) and $\eta$ is the fraction of raw supernova energy, $E_{\rm SN}$, that is converted into the energy gained by the gas. Assuming the supernova yield to be proportional to the mass of stars formed, $M_\star$, this can be written:
\eq
M_{\rm esc} \approx \left(\frac{v_{\rm SN}}{v_{\rm c}}\right)^2M_\star\hspace{2cm}\left[{v_{\rm SN}}^2\equiv\frac{\eta E_{\rm SN}}{M_\star}\right]
\qe
The original calculations by \scite{Larson74} proceed from an effective energy release per supernova of $10^{44}$J and a mass of stars formed per supernova of $100\Msun$, so the assumed yield of $10\%$ corresponds, in our notation, to $v_{\rm SN}^2\approx10^{41}{\rm J}/\Msun\approx (220\kms)^2$.

The adopted figure for energy release of $10^{44}$J by \scite{Larson74} was attributed to observational estimates by \scite{Minkowski67}, but has consistently appeared in the literature since then without any citation at all. More recent justification for this figure has been provided by fitting evolutionary models to the observed variation in luminosity of the supernovae 1987A (e.g. \pcite{Shigeyama90}), 1993J (e.g. \pcite{Nomoto93}), and 1994I (e.g. \pcite{Iwamoto94}), these having originated from stars with main sequence masses in the range 10-20$\Msun$, and thus being representative of the majority of the supernova population.

\begin{figure}
\includegraphics[trim =  8mm 52mm 74mm 78mm, clip, width=\columnwidth]{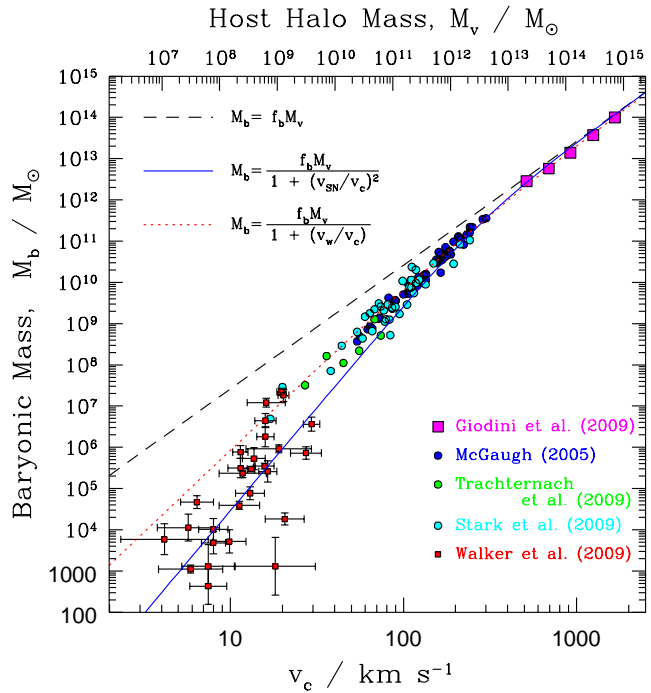}
\caption{Galactic baryonic mass estimates from several publications, shown as a function of the virialised host halo mass (estimated from observed circular velocity  or velocity dispersion). The lines show the simple prediction of self-similarity (dashed line), the prediction based on consistent conversion of supernovae energy to gravitational potential energy gain by the escaped gas (solid line and a similar prediction based on consistent momentum yield (dotted line) using the values $v_{\rm w} = v_{\rm SN}=300\kms$. This plot is deliberately similar in layout to fig. 1 of McGaugh et al. (2010) and uses the same conversion from characteristic speed to host halo mass: $M_{\rm v}/10^{12}\Msun \approx(v_{\rm c}/187\kms)^3$. Round symbols represent rotationally supported disks, where $v_c$ is the outermost measurement of circular velocity. Square symbols represent pressure supported systems for which $v_{\rm c}=\sqrt{3}\sigma$ is assumed, where $\sigma$ is the observed line-of-sight velocity dispersion.} 
\label{McGaugh}
\end{figure}

The scaling of eqn. \ref{Larson} was subsequently adopted by \scite{White78} as a component in their {\em analytic} model of galaxy formation, and the arguments behind it were revisited and extended by \scite{Dekel86}, though with rather detailed assumptions. \scite{White91} felt these were difficult to justify, but they did adopt the scaling (if not the parameter value) in their models. Indeed,  though different models employ their own particular definitions and contexts for its application, the assumption remains in use to this day (e.g. \pcite{DeLucia11}).

The mapping from supernovae physics to this observed trend in galaxy properties can be appreciated more directly, if more approximately, by simply combining eqn. \ref{Larson} with the additional constraint that the retained and escaped masses must add up to give the cosmic fraction, $f_{\rm b}$, of their host halo, ($M_{\rm b}+M_{\rm esc}=f_{\rm b}M_{\rm v}$). This leads to a simple relationship between remaining baryonic mass, $M_{\rm b}$, and total virial mass, $M_{\rm v}$ (or characteristic speed $v_{\rm c}$) that is plotted\footnote{The line in Fig. \ref{McGaugh} shows the locus of $f_\star\approx 1$. Real systems, where $f_\star\ltsim1$, would therefore be expected to lie above this line.} in Fig. \ref{McGaugh}:
\begin{equation}\label{fit}
M_{\rm b} \approx \frac{f_{\rm b}M_{\rm v}}{1 +f_\star\left(v_{\rm SN}/v_{\rm c}\right)^2}\hspace{2.2cm}\left[f_\star\equiv\frac{M_\star}{M_{\rm b}}\right] \label{ec}
\end{equation}
So, placed in the context of CDM cosmogony, \scite{Larson74} seems to have predicted, to an order of magnitude, the stellar masses for a wide range of cosmic structures and, crucially, this prediction stems from a simple physical premise.

But can basic energetic considerations really be a satisfactory explanation for this observed relationship? Can we really suppose that the converted fraction $\eta$ (or equivalently $v_{\rm SN}$) is the same in such a wide variety of environments? Even if so, why would we assume further that this remaining energy, carried by the out-flowing gas, is then almost exactly exhausted in overcoming the gravitational potential?

\subsection{Consistent momentum yield}\label{MomentumConservation}

The discussions reviewed so far have been focussed on the energy provided by supernovae. Yet, over time, stellar fusion generates a comparable amount of energy to the total produced by these more powerful, but rare events. The reason that supernovae are held responsible for the mass loss is because the energy released in this process couples effectively to baryons\footnote{The key comparison being between the opacity at the centre of a collapsing massive star, and that in a stellar atmosphere.}. With some fraction of the energy carried by massive particles, moving at some non-relativistc mean speed, $v_{\rm w}$, the {\em momentum} conveyed to the surrounding baryons per mass of stars formed ($p_{\rm SN}\sim E_{\rm SN}/v_{\rm w}$) can be much greater than (or at least comparable to, see \pcite{Murray05}) the total conveyed by the stars over their entire lifetime ($p_\star \ltsim E_\star/c$  where $c>>v_{\rm w}$). So it is the momentum yield that is crucial for driving a galactic wind.

An estimate of outward momentum may therefore seem a more relevant starting point for the derivation of the outflow mass. This is further motivated because this quantity is understood to be roughly conserved \cite{McKee77} after the initial adiabatic expansion phase. So the need to continue modeling the remnant through the cooling phase, to derive a final {\em energy} conversion, is eliminated from this alternative 1st-order estimate of the resulting outflow. 

Such an approach has some common ground with the one reviewed in \S\ref{EnergyConversion}, as {\em both} contain the same assumption of energy conservation\footnote{Both make use of an ``energy conservation'' argument, hence the need to take care when choosing terminology.} between some known phase in the galaxy, and some point far from the galaxy where all the energy is supposed to have been expended in overcoming gravity. In both cases the working assumption is that, in order to have escaped, the gas of mass, $M_{\rm out}$, must at some earlier point have been moving outward with velocity $v\gtsim v_{\rm c}$. 

The difference in this second case is to associate this with a {\em momentum}, of about $M_{\rm out}v_{\rm c}$, and suppose that this must (neglecting gravitational influence at this stage in the remnant's evolution) be less than or similar to the momentum yield at the end of the adiabatic phase. If it is this yield, per mass of stars formed, that is reasonably {\em consistent} across a range of systems ($p_{\rm SN} \approx M_\star v_{\rm w}$, where $v_{\rm w}$ is a mean system-averaged value for the specific momentum of the gas), this would imply:
\eq
M_{\rm out} \ltsim \left(\frac{v_{\rm w}}{v_{\rm c}}\right)M_\star~.\label{mom_cons}
\qe
The inequality also reflects the fact that this estimate of the ejected mass corresponds to its outward velocity tending to zero as it leaves the system (an approximation which will be redressed in \S\ref{Picture}). As can be seen in Fig. \ref{McGaugh}, this reveals a much tighter constraint on the low-mass outflow that was apparent from the energy budget. However, it is important to remember that in the {\em extreme} low mass limit there are no stars at all, and hence no supernovae. So perhaps we should not {\em expect} this asymptotic behavior to be fully accounted for by supernova winds, and should recognise the additional importance of natural cooling thresholds and gas stripping (e.g. \pcite{Stringer10}).

Though derivations differ considerably, this alternative scaling has become established as being associated with the momentum-conserving properties of winds (we emphasise again that this conservation is only one aspect of the argument) and has been applied when implementing supernova-driven winds (and indeed AGN-driven wind) in many published galaxy evolution models (e.g. \pcite{Oppenheimer06}).

\subsection{Application to phenomenological models} 

So far we have briefly reviewed two approaches which link escaped baryonic mass to the characteristic speed of the host system: the assumption of a consistent energy conversion of supernova energy to the outflow (eqn. \ref{Larson}) and the assumption that the momentum yield is consistent (eqn. \ref{mom_cons}). A third approach is to take a more general formula for the relationship, and tune the parameters to match simulations of the outflow process, and/or to match observations in conjunction with a galaxy formation model.

Such an approach was taken by \scite{Cole94}, who analysed the earliest {\em simulations} of galaxies by \scite{Navarro93}, where the effect of supernovae was explored by assigning velocity perturbations to the gas around star formation sites in their models. This implementation did not lead to the relationship given by (\ref{Larson}), but the results could be matched by the more general expression (in our notation):
\eq
M_{\rm out} \approx \left(\frac{v_{\rm SN}}{v_{\rm c}}\right)^{\alpha}M_\star~. \label{Cole}
\qe
where $M_{\rm out}$ is the mass of gas which has been ejected from the disk (as opposed to the gas which has escaped the host halo, $M_{\rm esc}$, used in eqn. \ref{Larson}.) This empirical approach is not generally physically motivated (though of course it reduces in the cases $\alpha=2$ and 1 to the two approximations described above).

The free parameter $\alpha$ was found to increase monotonically with the fraction of energy assigned to the velocity perturbations in the simulations (up to $\alpha=5.5$ in the case of a 20\% allocation).  This general relationship between mass outflow and circular (or escape) speed was then adopted in many models, often fitting $\alpha$ by empirical constraints, and it remains in use (for the latest discussion, see \pcite{Bower11}).

So, in the language of eqn. \ref{Cole}, $\alpha=1$, $\alpha=2$, and a whole range of other empirically motivated values are all being simultaneously called upon to account for various different observational features of the galaxy population. They can not all be correct at once. But arguing for any one simple approximation over the other is surely less progressive than seeking a more inclusive model which may {\em reduce} to certain simple scalings in particular regimes. Certainly, some alignment of these apparently conflicting working assumptions is essential if any kind of consensus on this aspect of galaxy formation is to emerge for the literature.
 
The first exercise towards this goal, presented in \S\ref{Results},  is to update the work of \scite{Navarro93} and \scite{Cole94} by investigating to what extent momentum and energy conversion are upheld by the results of the latest SPH simulations of the process. This study is integrated with an investigation into the modelling technique itself, considering variation of both wind parameters and the numbers of particles used to represent the systems. Guided by the implications of these numerical experiments, we then present in \S\ref{Picture} a revised analytic picture, which reflects the reality that gas will leave its host galaxy with a range of exit speeds, and includes both energy-conserving and momentum conserving constraints.

\section{Feedback in idealised disk galaxies}\label{Results}

\subsection{Summary of the numerical experiments}\label{Simulation}

The simulations of idealised, isolated disk galaxies that will be the focus of this section are generated from the same initial conditions as used by \scite{DallaVecchia08} (after \pcite{Springel05a}). We follow their evolution\footnote{This time interval was considered to be appropriate for the purposes of this study, being long enough to span many dynamical times for all the systems, whilst being short enough that the isolation from the cosmology was an acceptable approximation.}  for 0.5Gyr using the modified version of the {\sc gadget}-3 code (last described in \pcite{Springel05}).

This version has been used in the OverWhelmingly Large Simulations (OWLS) project \cite{Schaye10} and in the Galaxies-Intergalactic Medium Interaction Calculation (GIMIC, \pcite{Crain09}). In particular we use the `reference model' of OWLS. Briefly, sub-grid physics is implemented such that stars form at a rate which guarantees a Kennicut-Schmidt law for a disc in hydrostatic equilibrium \cite{Schaye08} and gas cooling includes the contribution from hydrogen, helium, 9  `heavy elements' and bremsstrahlung, as described in \cite{Wiersma09}. We assume the gas to have a solar abundance pattern with solar metallicity, and stars are assumed to form with the IMF proposed by \scite{Chabrier03}. Cosmological simulations using this code form plenty of galaxies that are similar to the Milky Way, in terms of stellar \cite{Font11} and X-ray properties \cite{Crain10}.
  
The idealised galaxies have dynamic baryonic components, with mass fraction 0.054, embedded in static dark matter halos containing the remainder of the total mass, $M_{\rm halo}$, with mass distribution  $M(r)=Mr^2/(r+r_{\rm Hq})^2$ \cite{Hernquist90}. The lengthscale, $r_{\rm Hq}$, for each galaxy is given in Table \ref{Details}. The baryonic mass is divided between a stellar bulge with total mass fraction of 0.014 (also distributed with a Hernquist profile) and a disk of fraction 0.04. 

Of the disk mass, 70\% is contained in a pre-existing disk of old stars\footnote{In this model, the contribution to the total energy yield from ongoing winds or supernovae from this ``old'' population is assumed to be negligable.} and the remaining 30\% is composed of cold disk gas. Both disk components have surface density distributions which decrease exponentially with radius with the same scalelength, set iteratively such that the angular momentum of the disk is a particular fraction of the total (with spin parameter, $\lambda=0.033$, see \pcite{Springel05}). The disk scaleheight and bulge scalelenghts are then both set to a tenth of this value, with the stellar disc following the profile of 
an isothermal sheet  and the vertical gas distribution set up in hydrostatic equilibrium using an iterative 
procedure. The simulations were run with a default gravitational softening length of 10pc, chosen so that this is always smaller than the Jeans length at the highest surface densities we expect\footnote{Note that since we integrate over a short time in a galaxy with a significant dark halo, two-body \em{relaxation} effects are mostly negligible.}.

\begin{table}\label{Details}
\footnotesize
\setlength{\tabcolsep}{2pt}
\setlength{\tabcolsep}{2pt}
\caption{Parameters of the idealised galaxy simulations.}
\begin{tabular}{@{}llccccc}
\hline
\multicolumn{2}{@{}l}{Total\,Mass/$\Msun$} & $10^{10}$ &$10^{10.5}$ & $10^{11}$ & $10^{11.5}$ & $10^{12}$\\
\hline
\multicolumn{2}{@{}l}{Galaxy\,Mass/$\Msun$} & $5.4\cdot10^{8}$ &$1.7\cdot10^{9}$&$5.4\cdot10^{9}$ &$1.7\cdot10^{10}$& $5.4\cdot10^{10}$\\
\hline
\multirow{2}{*}{Disk} 
&$r_{\rm D}$/kpc &0.53&0.78&1.1&1.7&2.5\\ 
&$z_{\rm D}$/kpc &0.053&0.078&0.11&0.17&0.25\\ 
\hline
{Bulge} & $r_{\rm Hq}$/kpc&0.053&0.078&0.11&0.17&0.25\\ 
\hline
Halo & $r_{\rm Hq}$/kpc &6.52&9.56&14.0&20.6&30.3\\ 
\hline
\end{tabular}
\end{table}

The effects of supernovae are simulated by assigning particles surrounding a star forming particle a velocity kick, of magnitude $v_{\rm w}$, in a random direction. Neighbouring particles are chosen at random such that their total mass is proportional to the initial mass of stars formed. The default choice for this ``mass loading'' in these models is:
\eq
\Delta M_{\rm w}=2\Delta M_\star~, \label{wind}
\qe
where $\Delta$ denotes a change over a given time interval. To yield an appropriate total energy, this mass loading corresponds to an initial wind kick, $v_{\rm w}=600\kms$ (giving $\Delta E_{\rm SN} \approx 7\times10^{41}(\Delta M_\star/M_\odot)$J). These fiducial values follow \scite{DallaVecchia08}, and are adopted here whenever other parameters in the model are under scrutiny. It is important to note that {\em all} the available supernova energy is allocated to these velocity perturbations, the radiation losses being left to emerge from the SPH calculation itself (rather than assumed upfront). Analysis of this emergent energy conversion is the focus of the following section, \S\ref{Preview}, which begins our analysis of these numerical experiments.

We note that this is by no means the only possible numerical experiment that can guide our understanding of these processes. By simulating only a small section of the interstellar medium, it is possible to resolve length scales which, in the implementation here, have to be left to the sub-grid configuration of the model. Such investigations are in progress \cite{Creasey12} but these smaller volumes must necessarily make other assumptions regarding their interaction with the system in which they are supposed to be imbedded. For this study, we choose to focus on simulations of entire galactic systems.

\subsection{Initial comparison with consistent energy conversion argumnets}\label{Preview} 

\begin{figure}
\includegraphics[trim =  10mm 110mm 102mm 35mm, clip, width=\columnwidth]{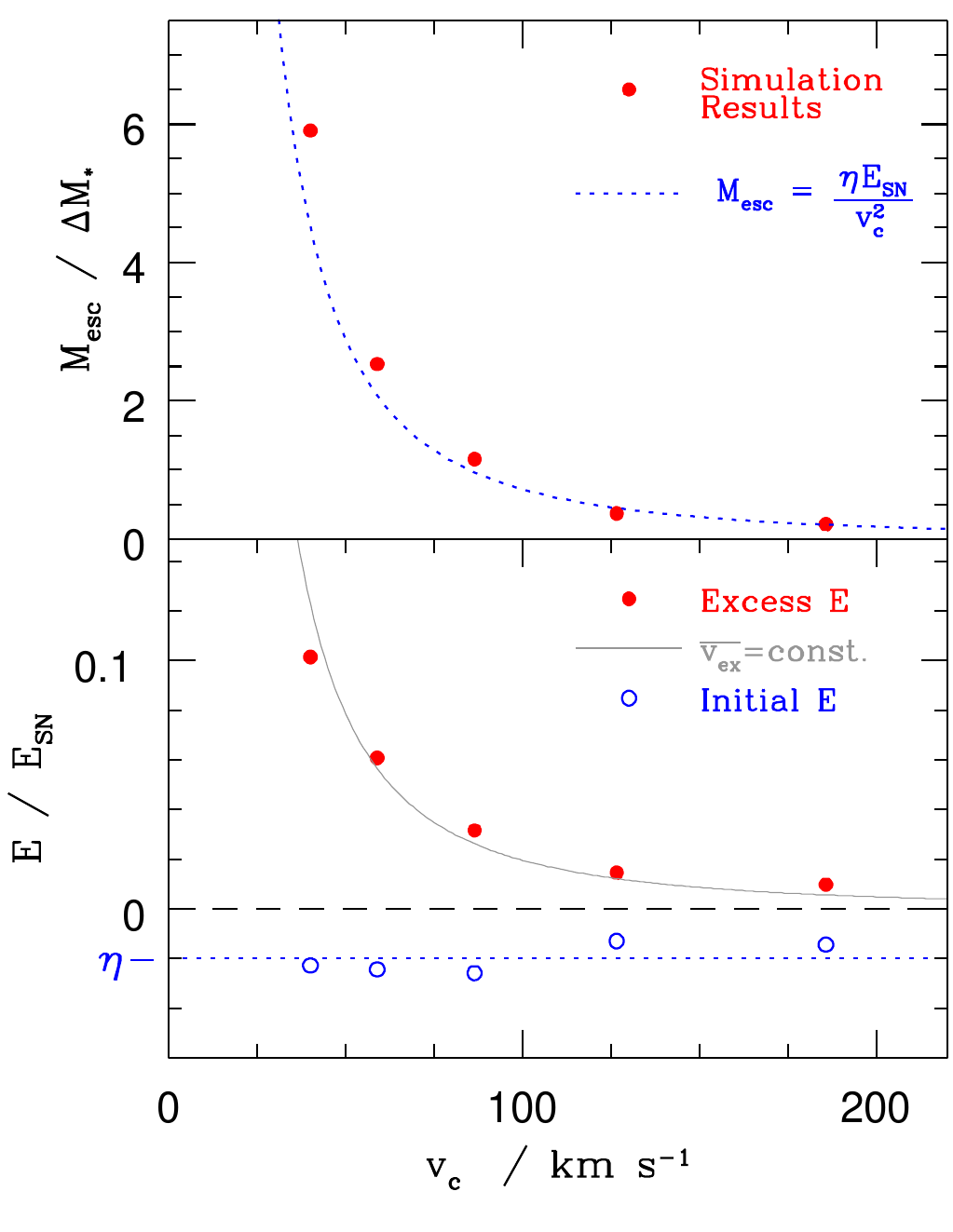}
\caption{A comparison between traditional feedback theory and results from state-of-the-art SPH simulations. The points in the {\bf upper panel} show the escaped gas mass, $ M_{\rm esc}$, in five idealised disk galaxies after 0.5Gyr of evolution (in units of the the mass of new stars formed, $\Delta M_\star$), as a function of their characteristic speed, $v_{\rm c}$. The galaxy masses range from $5\times10^{8}$ to $5\times10^{10}\Msun$. The {\bf lower panel} reveals the energetics. Solid points show the total energy that is escaping with the gas and open points show the initial energy of the escaped gas (after a few dynamical times, but before perturbation by supernovae). Both are plotted as a fraction of the total energy input. The {\bf dotted line} in both panels attempts to apply equation (\ref{Larson}) to these results. In the bottom panel shows the mean of the measured efficiencies, $\eta$. The top panel puts this value back into (\ref{Larson})  to see if it agrees with the results. The faint solid line in the lower panel shows the approximate outgoing energy which would result from the mass scaling in this equation, together with a common mean exit speed$^4$, $\overline{v_{ex}}$. }\label{escape}
\end{figure}

To connect with the review of analytic treatment  in \S\ref{Introduction}, we begin by presenting a few pertinent results from five idealised disk galaxies covering a range of host halo (and galaxy) mass. The results are plotted in Fig. \ref{escape} as a function of characteristic velocity, $v_c$, here defined  as the root-mean absolute specific energy, $v_{\rm c}\equiv\sqrt{-\bar{e}_0}$, measured for all the gas in the galaxy (after several dynamical times, but before perturbation by supernovae). This working definition is adopted as it is a well-defined quantity in the simulation and is consistent with its  application in the analytic model (which is concerned with the energy deficit that has to be overcome). Also, this and the observable $v_{\rm c}$ are mutually satisfactory proxies.

The top panel shows the mass, $M_{\rm esc}$, which has ``escaped'' the galaxy (outside the disk and with positive total energy). The lower panel shows the total\footnote{ Kinetic, potential and thermal energy are all included.} energy of this body of gas, the solid points being the value at the final time-step.  To provide a context for these ``excess'' energies, a faint line is included showing the approximate scaling that would result from a common residual wind velocity\footnote{Direct measurements of this reveal true mean values in the range $\overline{v_{\rm ex}}\sim 100$ to $150\kms$, increasing with $v_{\rm c}$. Exit velocities are investigated in \S\ref{Convergence}.}, $\overline{v_{\rm ex}}=140\kms$, and the mass scaling of equation (\ref{Larson}).

For now, the most important thing to note is that these measured residuals of {\em final} energy are just as significant, if not more so, than the {\em initial} energy of the same body of gas, shown as open points in the lower panel.  So in this particular model, at least, it seems that there is much more energy expended than is required to just ``remove gas from the galaxy'', as described in the quotation in \S\ref{EnergyConversion}. A significant fraction is expended in the form of kinetic and thermal energy of the gas as it escapes. In the case of less massive galaxies (where the wind velocity turns out to be several times greater than the characteristic velocity) it is actually the gravitational energy that could be neglected, if anything.

This can be clarified by reference to an extended version of (\ref{Larson}) which identifies some other energy sinks:
\eq
E_{\rm SN}  = E_{\rm lost} + \frac{1}{2}M_{\rm esc}v_{\rm c}^2 + E_{\rm trapped}+ E_{\rm ex}\label{old}
\qe
Equation (\ref{Larson}) implicitly assumes that the final two terms here can be neglected. As just discussed, these particular results clearly lend no support for neglecting the excess energy, $E_{\rm ex}$.  Indeed, in reality, it is difficult to see how the mass of gas affected by the supernovae could be naturally tuned such that it will always coast gently out to rest in the intergalactic medium.

Conversely, it also seems probable that plenty of gas will acquire some energy from supernovae but not {\em enough} to escape (corresponding in eqn. \ref{old} to $E_{\rm trapped}>0$). In order to satisfy energy balance of the system over many dynamical times, there is an argument that this must be eventually lost (and hence covered by $E_{\rm lost}$). But, even if so, this ``fountain'' process must still be considered in any reasonable derivation of the total energy loss.

Yet, despite all these caveats, the measured mass outflow in the model, shown in the upper panel of Fig. \ref{escape}, does appear to agree with (\ref{Larson}).  The reason for this can be quickly appreciated: Though the {\em total} energy fraction converted is variable (as discussed above), the {\em potential} energy which has been overcome to escape is similar in all systems. So the assumption that it is the {\em only} energy sink, though in fact incorrect, appears to work.

This is shown by the open points in Fig. \ref{escape}, together with a dotted line indicating the mean value, $\eta$. When this value\footnote{In this case $\eta= 2.4\%$ and $E_{\rm SN}=(600\kms)^2M_*$, so $v_{\rm SN}=93\kms$.} is used in (\ref{Larson}), the small scatter around it means that there is reasonable agreement with the measured {\em mass} in the upper panel.

So here we have a plausible scenario in which a frequently adopted scaling {\em does}, broadly, apply, but in which the usual assumptions are found to be only {\em partially} true. This is an illustration of the dangers of confirmation by affirmation. However, an {\em illustration} is all that it can be at this level of analysis. The degree to which any particular implementation of winds, or set of numerical methods may represent reality is far from established. What we {\em can} try to establish is this:
\begin{itemize}
\item{The physical cause for certain aspects of the model's behaviour.}
\item{Which aspects therefore seem unreliable?}
\item{Which aspects are indeed telling us about real galaxies?}
\item{How can our physical understanding of the process be modified in the light of this investigation?}
\end{itemize}

\begin{figure}
\includegraphics[trim =  96mm 95mm 17mm 25mm, clip, width=\columnwidth]{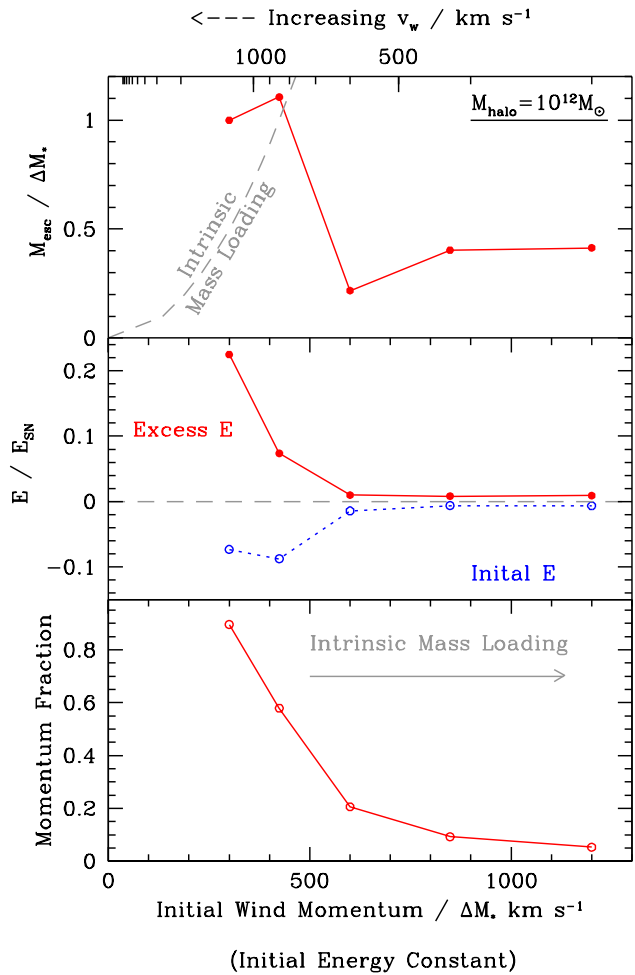}
\caption{The effect of assuming different intrinsic mass loading in the modelling of supernova-generated wind, given a fixed energy yield for a halo of $M_{\rm halo}=10^{12}\Msun$. As the increased mass loading leads to a higher initial momentum, this quantity is plotted as the main x-axis (the wind speed decreases from left to right, as shown at the top of the figure). The top two panels show the same quantity as those of Fig.\ref{escape}. The bottom panel shows the momentum of the gas out of the plane of the galaxy, as a fraction of the initial momentum in these directions.}\label{ml}
\end{figure}

\subsection{Variation of injected wind momentum }

To begin analysing the modelling method, we briefly present the effect on the outflow of varying the wind speed and mass loading. To help isolate the effect, the total wind {\em energy} input,  $E_{\rm SN} = \frac{1}{2}M_{\rm w}v_{\rm w}^2$, is kept constant for all the realisations in this test.

Fig. \ref{ml} shows the mass, energy and momentum out of the plane that escape the same galaxy under five different initial wind speeds, but with the same total energy budget. With high wind speed, to the left of the plot, a small number of gas particles (about ten thousand per timestep) receive a large velocity kick. At these high speeds, the change in speed due to gravity is small and almost all of the momentum out of the plane is conserved in the outflow.

With low wind speed, towards the right, a larger number of particles (a few hundred per timestep) receive a small kick, and almost all the energy and momentum is absorbed. However, both the energy and, crucially, the {\em mass} which escapes are quite robust to changes in mass loading at the highest values of this parameter. It should be noted that the default choice ($v_{\rm w}=600\kms$) is in the range where the mass outflow is sensitive to this value, for this halo mass.

\subsection{Convergence and limiting behaviour}\label{Convergence}
 
In addition to the importance of parameter choice, galaxy simulations such as these can also be very sensitive to resolution. In this section, we illustrate the differing effects of {\em the same} physical assumptions applied to initially identical model galaxies which are resolved at a range of {\em different} values for gas particle mass, $m_{\rm p}$. 

For a clear study of the effects of varying particle mass, the same gravitational softening length was used for all the runs.  Different values would have been adopted for different resolutions if the simulations were being run in their own right, rather than as part of a convergence test. A different choice would have had a lesser effect than varying the particle mass (see Appendix \ref{Softening}).

For this more detailed investigation, we adopt a slightly more precise definition of ejected gas than was used in \S\ref{Introduction}. We define two planes which initially enclose all the particles at the highest resolution\footnote{The highest resolution was limited by the number of particles (rather than particle mass), and a practical maximum number turned out to be, for gas particles, $N_{\rm p}=2.3\times 10^5$.} by a margin of 10\%. (So the outflow boundary exceeds the unperturbed height of any particle by a factor of at least 1.1).  It turns out that this is almost the same for all systems, between 2 and 2.2kpc above and below the disk plane. A number of useful quantities can then be measured from each simulation, for any chosen time period up to the total duration of the calculation:

\begin{itemize}
\item{The gas mass which flows out through these two boundaries, $M_{\rm out}$.}
\item{The distribution of this mass as a function of velocity component perpendicular to these planes, $v_z$.}
\item{The mass of gas which flows back through the boundary, having been outside it (``re-cooled'')}
\end{itemize}

For all of these components, it is instructive to distinguish between ``wind particles'', that were directly assigned kinetic energy as a result of being near a simulated supernova, and particles that have been indirectly affected.  The total mass of directly effected wind particles is denoted $M_{\rm wind}$. From the geometry of the randomly orientated kicks that they receive in this implimentation, the initial distribution of velocities out of the plane is ${\dif M}/{\dif |v_z|} = {M_{\rm w}}/{v_{\rm w}}$ and the initial momentum away from and perpendicular to the plane is $\frac{1}{2}M_{\rm w}v_{\rm w}$. 

\subsubsection{Dwarf galaxy}\label{Dwarf}

\begin{figure*}
\vspace{5mm}
\subfigure{\includegraphics[trim =  140mm 50mm 140mm 30mm, clip, width=\columnwidth]{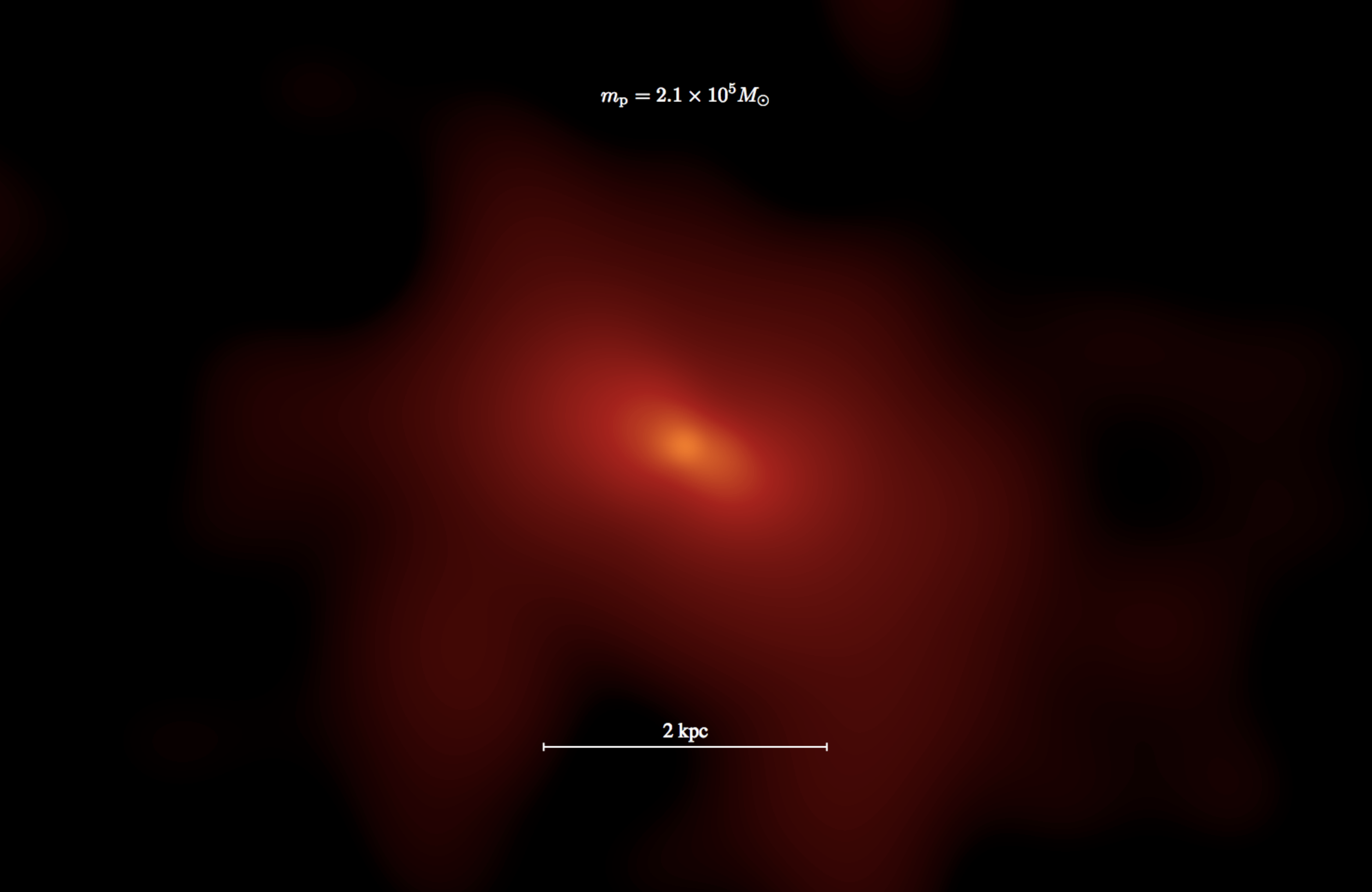}}
\hspace{6mm}
\subfigure{\includegraphics[trim =  140mm 50mm 140mm 30mm, clip, width=\columnwidth]{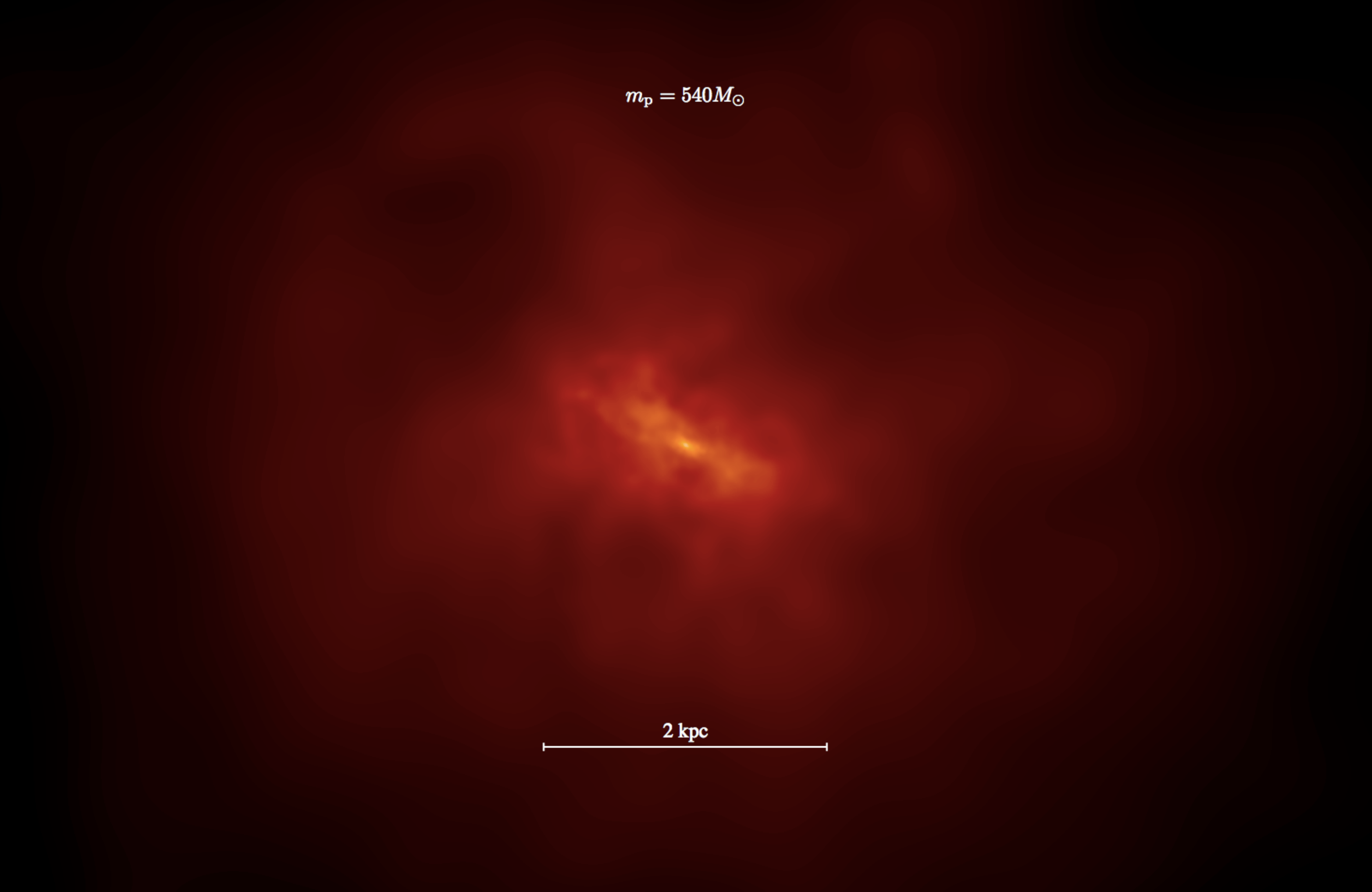}}
\caption{Two realisations of the $10^{10}\Msun$ halo, showing the projected gas density (brighter regions show higher densities). The left panel shows a low resolution version  (which starts with 560 gas particles), corresponding to the second from left column in Fig. \ref{m10}. The right panel is the highest resolution that was achievable (750,000 gas particles), corresponding to the far right column in Fig. \ref{m10}. Both show the system at the final timestep, after 0.5Gyr of evolution.}\label{images10}
\vspace{1cm}
\end{figure*}

\begin{figure*}
\includegraphics[trim =  7mm 55mm 17mm 25mm, clip, width=\textwidth]{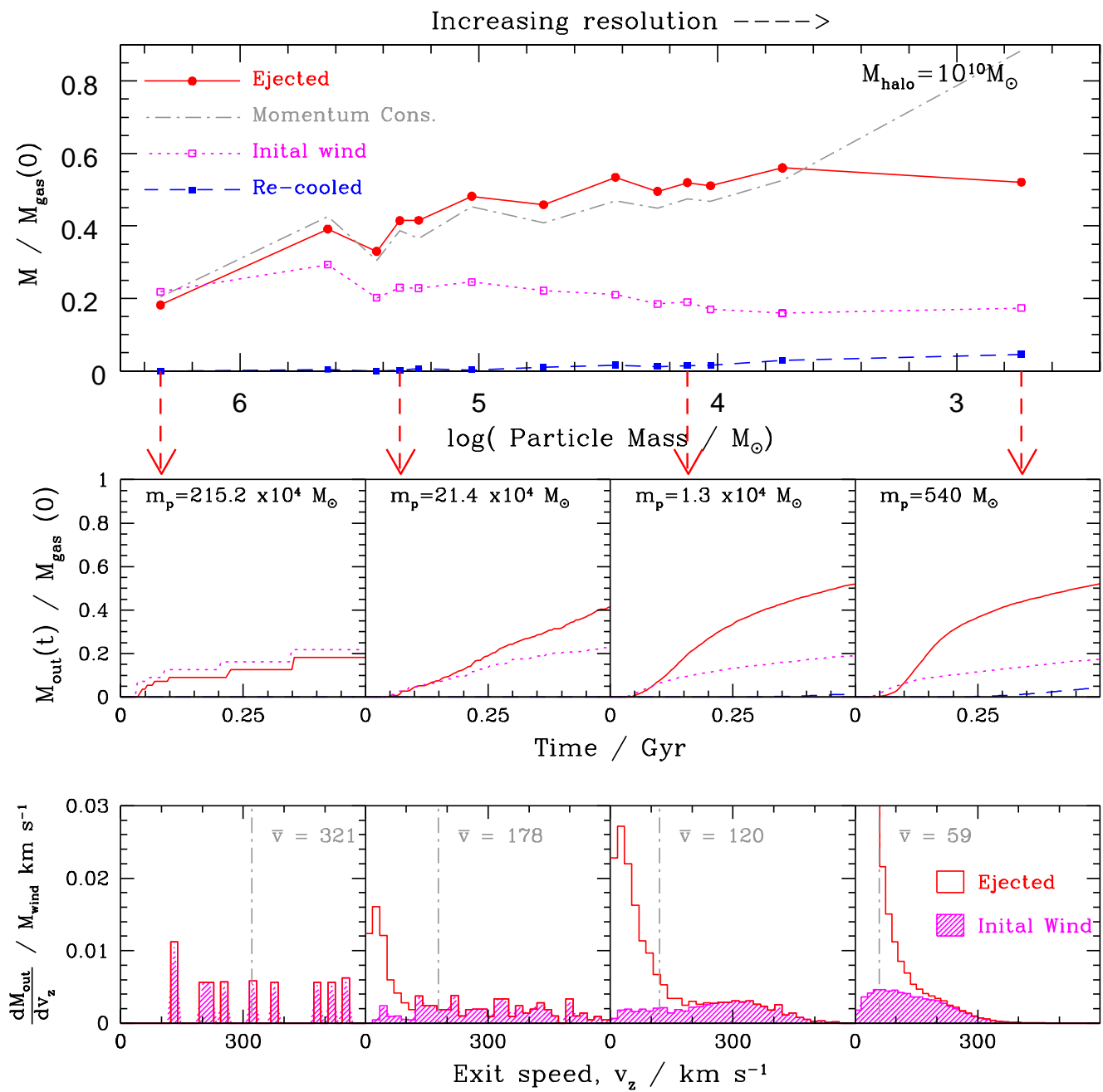}
\caption{The effect of resolution on the mass outflow and evolution of an idealised, isolated dwarf galaxy (halo mass$=10^{10}\Msun$, baryonic fraction=0.054). The {\em same} initial conditions have been evolved using the {\sc gadget} code but with 15 different choices of particle mass. The gravitational softening length in all cases is 10pc. The {\bf middle panels} show the time evolution of four of these, showing the gas mass fraction which has escaped (solid line), and the fraction which has escaped from the disk but returned (dashed line). Also shown is the fraction which has been directly allocated wind energy (dotted line), which in this set of simulations is just $M_{\rm w}=2\dot{M}_\star$. The {\bf upper panel} shows the final value of each of these quantities as a function of the particle mass used in the simulation run. Added to this panel is a faint dot-dashed line showing the result of straightforward conservation of initial momentum out of the plane (\ref{momentum}). The {\bf lower panel} shows the distribution of escaped gas as a function of the outward velocity component, $v_z$, at the point when it crosses the nominal galaxy ``boundary'' (see main text). The vertical dot-dashed line shows the mass-weighted mean, $\overline{v_z}$ (in $\kms$) which is used in eqn. \ref{momentum}. Note that the integral under the total ejected mass is equal to $M_{\rm out}/M_{\rm wind}$, but the shaded area (which includes only wind particles) is not equal to unity because some wind particles do not escape.}
\label{m10}
\end{figure*}

Firstly we focus on a dwarf galaxy with $M_{\rm halo}=10^{10}\Msun$, where the effects of using different resolutions are particularly palpable. Images of the gas in the simulation at two different resolutions are shown in Fig. \ref{images10} and some quantitative results from simulations of this system, across a range of resolutions, are shown in Fig. \ref{m10}. The top panel of this figure shows the total gas mass which has been directly accelerated by simulated supernovae (the ``initial wind''), the total mass which has been ejected beyond the nominal galaxy boundary (defined above) and also the re-cooled mass which has returned to the inner region after being ejected.  

These cumulative values, over 0.5Gyr of evolution, provide a simple illustration of the behaviour of the simulation, which changes markedly as more and more particles are used to represent the galaxy. At the very lowest resolution, all the supernovae in the galaxy during a timestep are represented by  a {\em single} gas particle which is allocated all the energy and duly exits the disk. So in this limit, $M_{\rm out}\rightarrow M_{\rm wind}$. 

\begin{figure*}
\subfigure{\includegraphics[trim =  140mm 50mm 140mm 30mm, clip, width=\columnwidth]{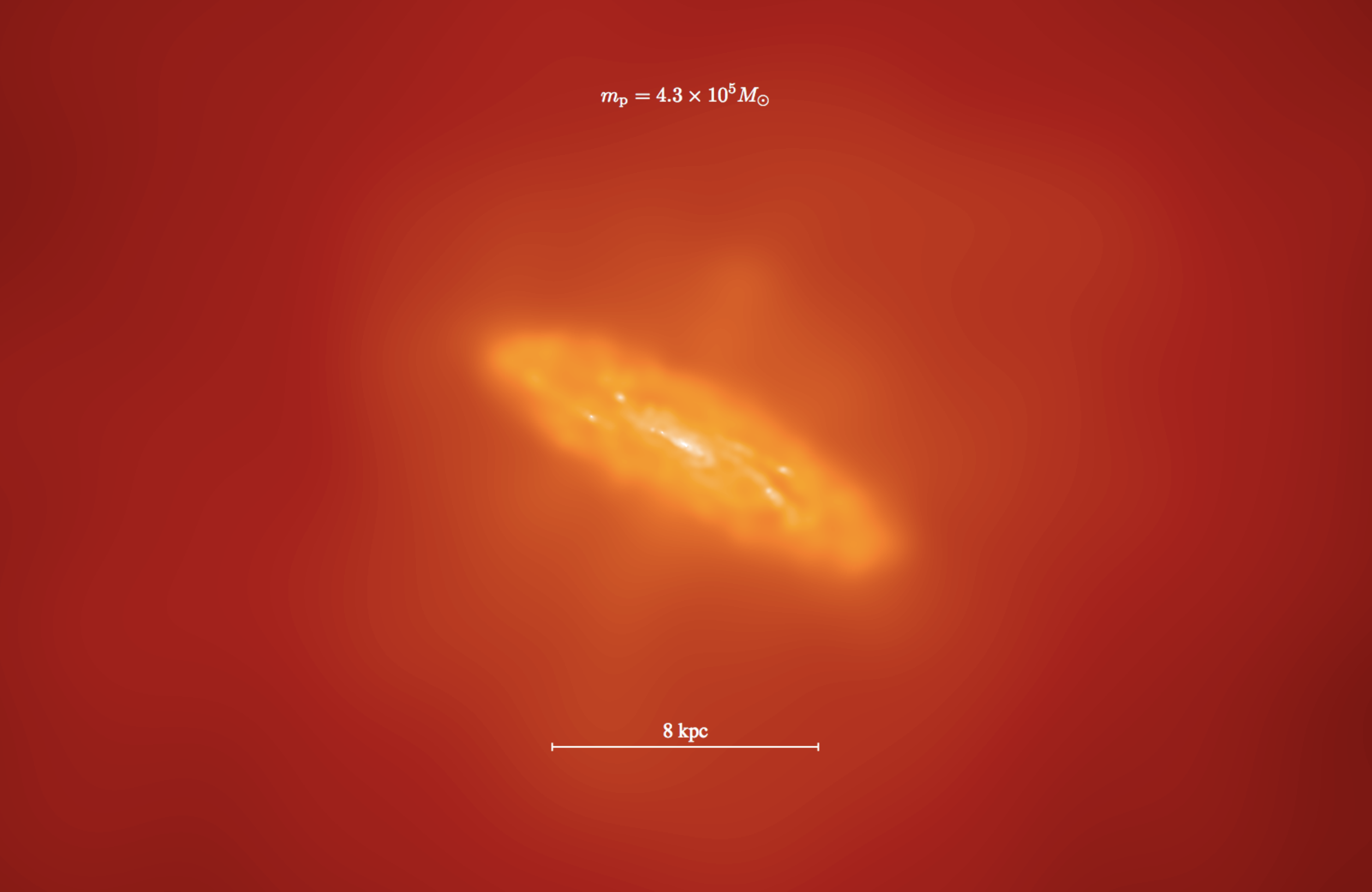}}
\hspace{0.6cm}
\subfigure{\includegraphics[trim =  140mm 50mm 140mm 30mm, clip, width=\columnwidth]{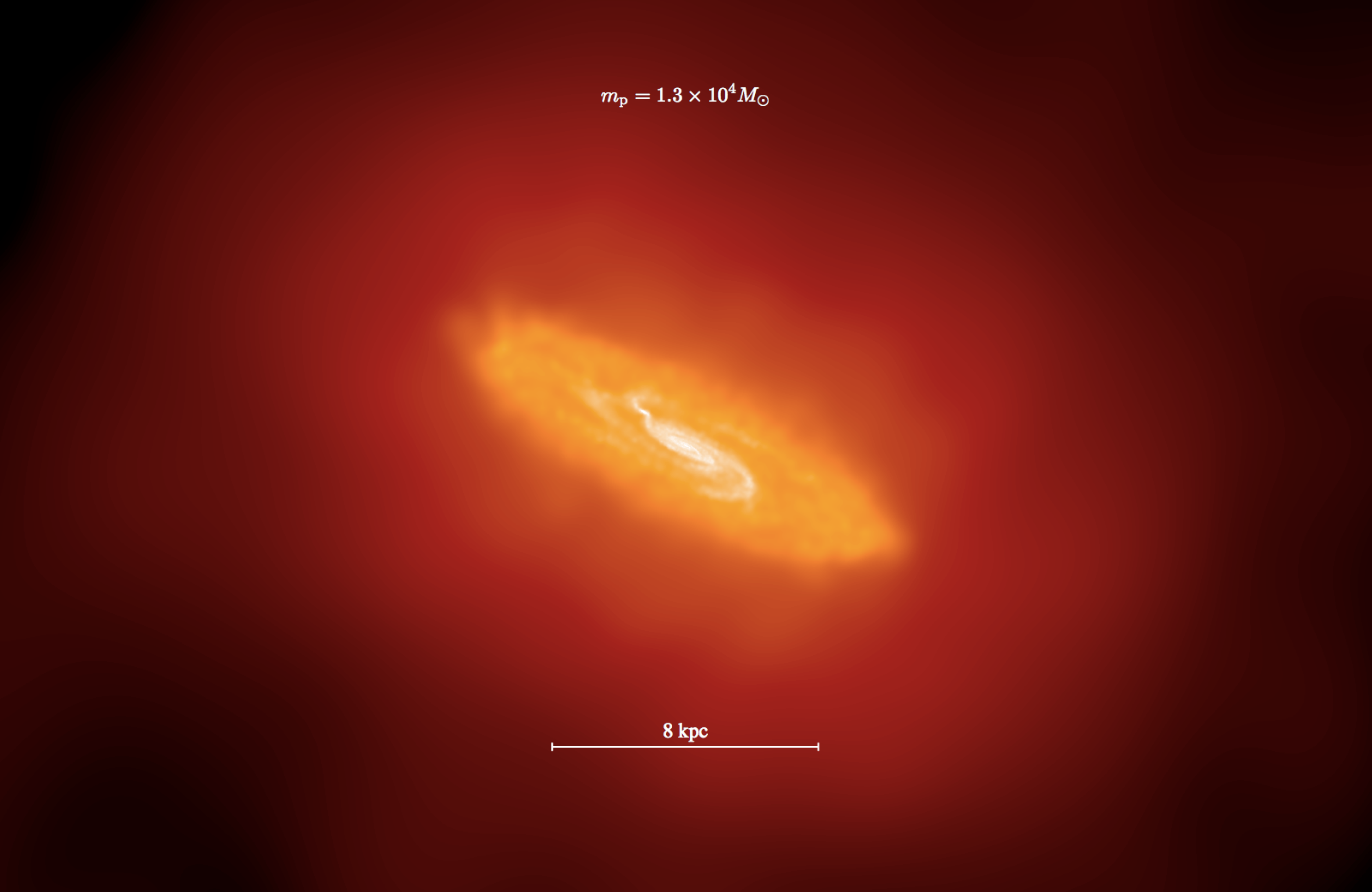}}
\caption{Two realisations of the $10^{12}\Msun$ halo, in the same format as Fig. \ref{images10}, but at a larger scale (as shown). The selected resolutions correspond to the second column (93,000 gas particles) and fourth column (3 million particles) in Fig. \ref{m12}. In this figure, and in Fig.\ref{images10}, each particle is represented by a smooth radial density profile which reduces to zero at a distance of $2s$, where $s$ is the lesser of a stipulated maximum, $s_{max}$, and the distance to the 16th nearest neighbour  ($s_{\rm max} = [1.5, 6.5]$kpc for $M_{\rm halo} = [10^{10},10^{12}]\Msun$).}\label{images12}
\end{figure*}

\begin{figure*}
\includegraphics[trim =  7mm 55mm 17mm 25mm, clip, width=\textwidth]{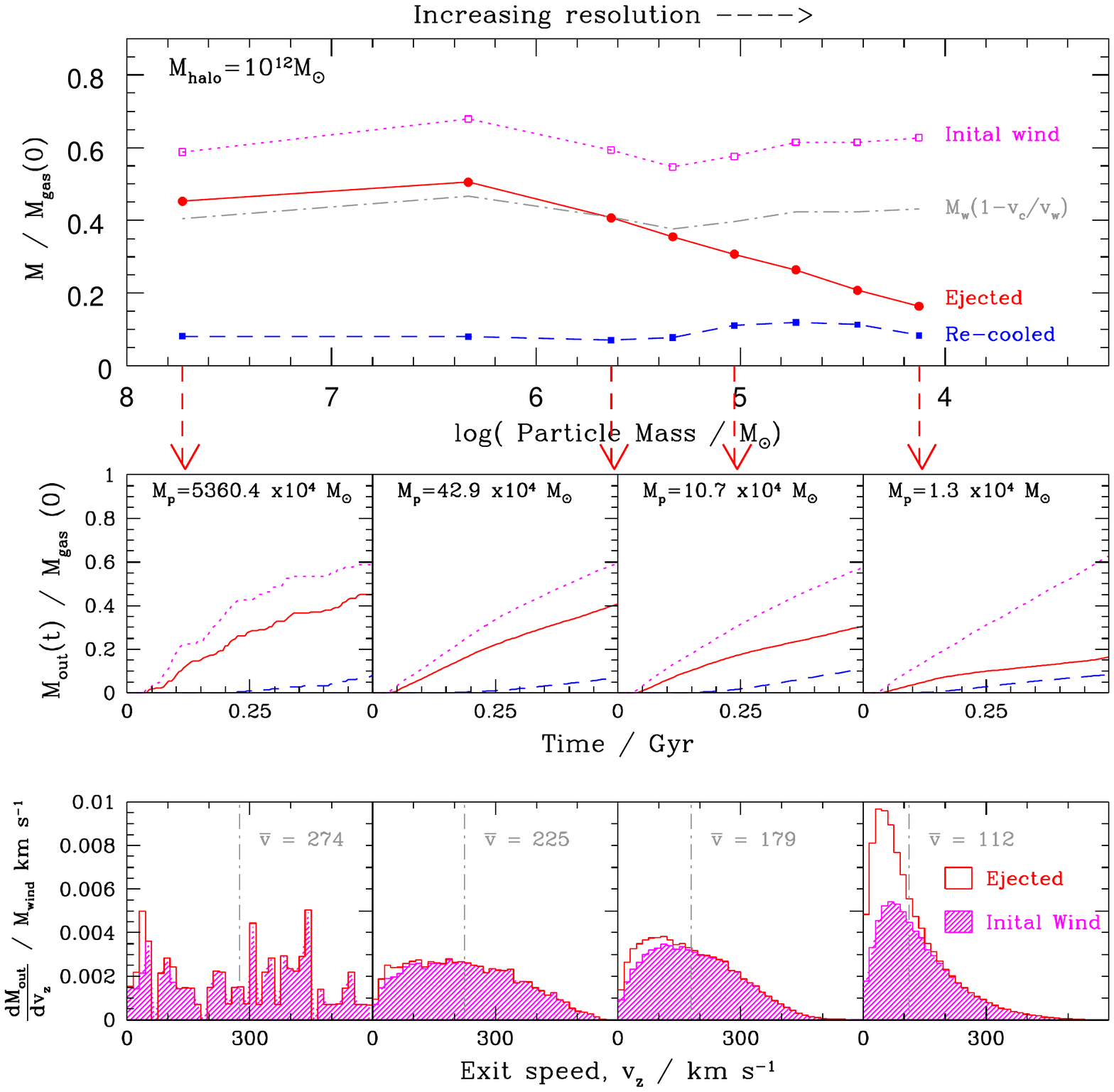}
\caption{The effect of resolution on the mass outflow and evolution of an idealised, isolated Milky Way-like galaxy (halo mass$=10^{12}\Msun$, baryonic fraction=0.054). The key is the same as Fig. \ref{m10}, except that the faint dot-dashed line for the limiting behaviour is different, as explained in \S\ref{MW}. Note, again, that the mass of stars formed is just half the mass of the initial wind (dotted line).}\label{m12}
\vspace{5cm}
\end{figure*}

This can be appreciated by reference to the middle row in Fig. \ref{m10}, which shows the time evolution of three of the cumulative quantities from the top panel. In the leftmost panel, the ejected mass fraction can be seen to be increasing in steps, corresponding to single wind particles exiting the disk region\footnote{The difference between ejected and wind mass in this case is due to a single particle which acquired an energy kick at $t\sim0.1$Gyr but remained contained in the disk. All the others escaped, so the two lines differ by the mass of one particle. The one particle happened to be contained because of its initial trajectory towards the disk centre.}. Gas which is not directly kicked by supernovae acquires very little of the energy and does not escape the disk.

With more particles, and more interactions, such secondary outflow does become possible. Because of this, the  evolution of the system looks very different to the right of Fig. \ref{m10} compared to the left. As the number of particles increases (left to right), the initial supernova momentum is shared by a greater gas mass. So the total {\em energy} which the outflow carries out is less, but the mass loss is greater. To accompany this explanation,  a faint line is included in the upper panel of Fig. \ref{m10} which shows the momentum conserving case:
\ar
M_{\rm out} &\approx& \frac{p_{\rm SN}}{\overline{v_z}} \nonumber\\
&\approx& \frac{\frac{1}{2}M_wv_w}{\overline{v_z}}, \label{momentum}
\ra
where $\overline{v_z}$ is the mass-weighted mean of the exit velocities. 

At low and intermediate resolution, interactions with the remaining disk gas are not generally sufficient to bring gravitational deceleration into play (i.e. $v_z^2>>v_c^2$ for much of the ejected gas) so equation (\ref{momentum}), which neglects gravitational influence, is a good fit to the results in this regime. As the resolution is increased and exit speeds are reduced, more and more mass is ejected from the system. Eventually, at the highest resolutions, the speeds are reduced to $v_z\ltsim v_{\rm c}$, gravity has a significant influence, and the results begin to diverge from the simple prediction of (\ref{momentum}). 
 
So, the broad behaviour of this numerical experiment can be understood in terms of physical arguments, and the principal results -- star formation and outflow -- do appear to be reasonably robust to changes in resolution above some reasonable particle mass ($\sim 10^4\Msun$). This is supported by the behaviour shown in the top panel of Fig. \ref{m10}. However, though the results may be {\em consistent} over this high-end range of resolution, the changing velocity distribution in the bottom row indicates that the calculation has not fully {\em converged} on all scales.

\subsubsection{Milky Way-like galaxy}\label{MW}

We now turn to the case of a larger galaxy, with $M_{\rm halo}=10^{12}\Msun$. Images of two realisations of the system can be seen in Fig. \ref{images12}, and the key results are shown in Fig. \ref{m12} (in the same format as Fig. \ref{m10}). 

To understand the behaviour of this larger system, we begin with the low resolution limit where interactions between wind particles and the rest of the disk gas are negligible. The outcome is nearly as straightforward as the low resolution limit in the dwarf galaxy, where almost all wind particles simply escape. But because this larger system has at least some gravitational influence on the wind particles ($v_{\rm c} \sim v_{\rm w}$) a significant fraction are retained even in the absence of significant additional interaction. This statement implies something along the lines of:
\eq
M_{\rm out}\approx \int_{v_{\rm cut}}^{v_{\rm w}}\frac{\dif M}{\dif v_z}\dif v_z ~. \label{integral}
\qe
For the simple low-resolution case of a flat initial mass distribution  in $v_z$, which results from randomly directed velocity kicks, equation (\ref{integral}) becomes:
\begin{equation}
\frac{M_{\rm out}}{M_{\rm w}} \approx 1 - \frac{v_{\rm cut}}{v_{\rm w}}.\hspace{1.5cm}\left[{\rm for}~~\frac{\dif M}{\dif v_z}=\frac{M_w}{v_w}\right]\label{flat}
\end{equation}
A dot-dashed line is included in Fig. \ref{m12} to illustrate this relation, using $v_{\rm cut}\approx v_{\rm c}$ as a nominal value for the velocity required, on average, for gas to escape.

Of course, the distribution of wind velocities is only flat at the lowest resolutions, so this does not match the results of the high resolution runs. As the system is modelled with more and more particles, the velocity distribution tends to the smooth form that can be seen in the bottom right panel. Notably, at the very highest resolution, the distributions of directly and indirectly affected particles are beginning to look quite similar. This is an encouraging sign that the calculation is beginning to converge, but the total mass outflow is still lower with each successive increase in particle number.

Concern over the lack of convergence in Figs. \ref{m10} and \ref{m12} could be compounded by concerns for the particular implementation of winds that has been used in the models. Idealistically, one could argue that the precise choice of implementation would not matter with a sufficiently reliable SPH calculation, ejecting energy as random velocity perturbations to the particles being in principle equivalent to updating the ``thermal'' energy of each particle by the appropriate value. But, in practice, such concordance is not the case without a more efficient time-integration scheme, as shown by \scite{Durier12}. 

Some simple tests, such as shown in Appendix \ref{Timesteps}, suggest that the consequences for these particular experiments (even in the case of the low resolution runs) is fortunately not critical. But ultimately, this is yet another reason to remember that the results presented in this section are an {\em investigation} rather than a test. That said, we return to our premise at the end of \S\ref{Preview}: If we can at least understand the behavior of our {\em models} (incomplete though they may be) in a physical way, we can use this to help motivate a more comprehensive understanding of {\em real} systems (even though the model itself may not perfectly represent them).

Thus, if  (\ref{integral}) works well for the simple velocity distributions seen in Fig. \ref{m12} at low resolution, it might well be expected still to apply at the highest available resolution {\em and indeed in reality}, if only we knew the appropriate velocity distribution to use. So, if we could anticipate the velocity distribution towards which the simulated winds are beginning to converge, we would equally well be able to predict - and understand - the mass outflow that would emerge at arbitrarily high resolution. This prospect will be examined in \S\ref{Picture}.

\subsection{Mass dependence}\label{Mass}

\begin{figure}
\includegraphics[trim =  95mm 93mm 17mm 35mm, clip, width=\columnwidth]{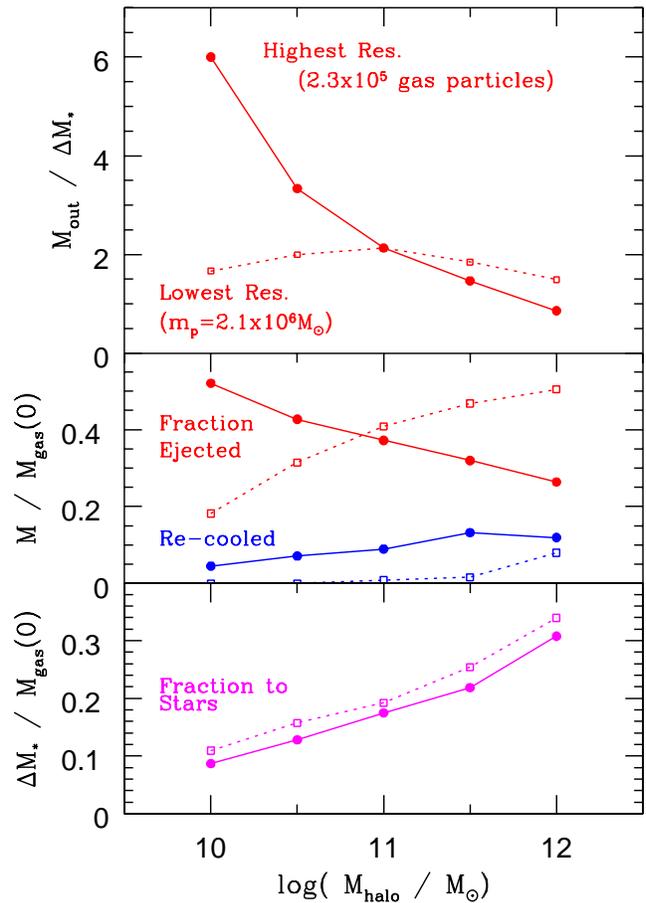}
\caption{The outflow of gas from idealised galaxies of different total mass, $M_{\rm halo}$. The {\bf top panel} plots the ratio between the ejected mass and the mass of stars formed. The {\bf middle panel} shows the fractions which have been ejected from the system and re-cooled and the {\bf lower panel} shows the fraction of the initial gas mass that has formed stars after being evolved for 0.5Gyr (circles). In all panels, the highest resolution is shown as solid lines/points and the lowest choice as dashed lines and open points.}\label{beta}
\end{figure}

The goal set up by \S\ref{Introduction} was to acquire a better understanding of the correlation between outflow efficiency and host galaxy properties, in particular the mass (or circular velocity). To reinitiate this discussion, Fig. \ref{beta} presents some key results from the idealised simulations as a function of host halo mass, but now showing not only the highest available resolution that was attained for all systems (using the same number of particles), but also the equivalent results at very low resolution. Reassuringly, the star formation appears to be robust to variation in resolution, so this discussion can focus on the behaviour of the outflow

Turning to the outflow mass, the lower resolution case is easily understood by noting that, at all halo masses, the outflow mass is approximately double the mass of stars formed (clear from the upper panel), which is simply the ``mass loading'' employed in this feedback prescription (\ref{wind}). Broadly, all kicked particles escape, others are unaffected.

At the highest possible resolution, the mass outflow is highly dependent on the local environment. As discussed in \S\ref{Convergence}, this variation is the result of two principal competing effects:
\begin{itemize} 
\item{the distribution of energy across a greater gas mass \\
(tending to increase the ejected mass due to conservation of momentum) and} 
\item{the radiation of energy that such distribution will produce \\
(tending to reduce the ejected mass; lower velocities are insufficient to escape gravity).}
\end{itemize} 
A quantitative account of the relative importance of these two influences will be the focus in \S\ref{Picture}.

\section{Towards an improved picture}\label{Picture}

Armed with a better understanding of the simulations, we return to the task motivated in \S\ref{Preview}: to seek a link between the {\em external} effect of supernova feedback (e.g. $M_{\rm out}$), the {\em internal} conditions in the galaxy (e.g. $\overline{v_z}$)  and the governing parameters (e.g. $v_{\rm w}$). Some steps towards this have already been taken in limiting cases.

The case of a neglible potential barrier, but significant energy losses, was satisfactorily accounted for by a momentum conserving argument (\ref{momentum}). This was compatible with the outflow in the limiting case ($v_{\rm w}^2>>v_{\rm c}^2$) a good approximation for the lowest mass galaxy up to quite high resolution (Fig. \ref{m10}) and also for the most massive galaxy when simulated with a very high choice of intrinsic wind speed (Fig. \ref{ml}).

The case of a significant potential barrier, but negligible energy losses, was successfully accounted for by   (\ref{integral}). When applied to the limiting case of ballistic wind particles, perturbed only by gravity,  (\ref{flat}), this accounted for the low-resolution limit in the high mass galaxy (Fig. \ref{m12}).

\begin{figure}
\subfigure{\includegraphics[trim = 96mm 55mm 16mm 76mm, clip, width=\columnwidth]{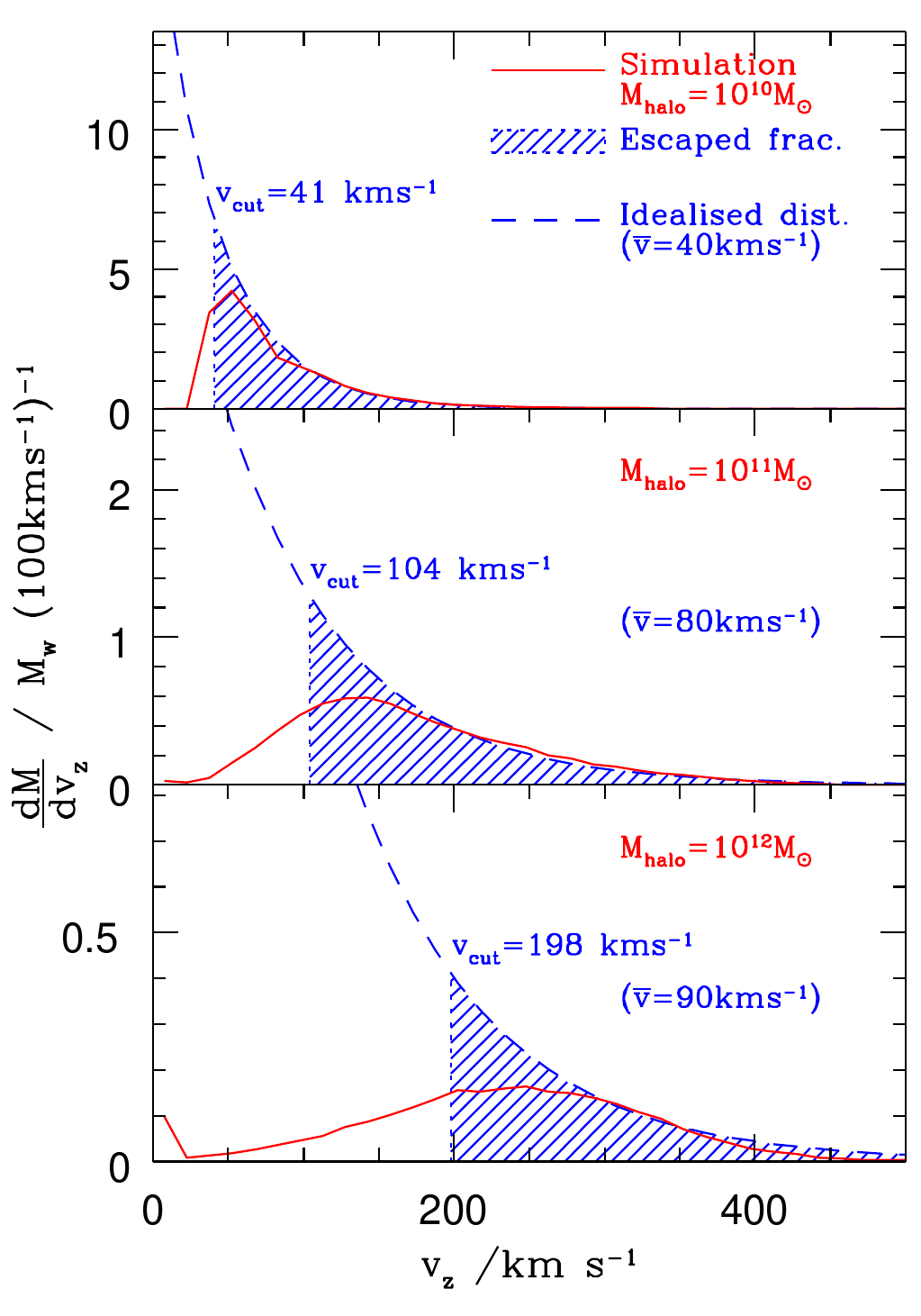}}
\subfigure {\includegraphics[trim = 96mm 170mm 16mm 36mm, clip, width=\columnwidth]{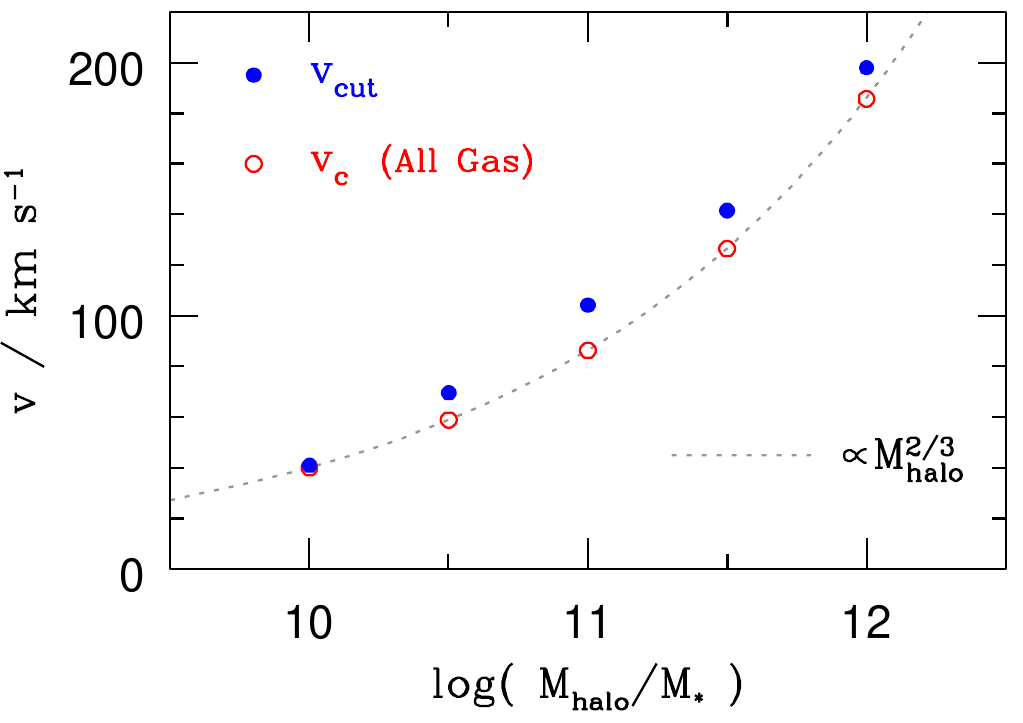}}
\caption{A comparison between the approach to modelling outflow described by equation (\ref{general}) and the results of the controlled simulations of disk galaxies, run with the same number of particles ($\approx 2\times 10^5$ gas particles).  The {\bf upper panels} show distribution of ejected gas as a function of outward velocity, from the simulation results (solid line) and according to eqn. \ref{dMdv} (dashed line). The mean, $\bar{v}$, of the velocity distribution (\ref{dMdv}) is fitted to match the simulation results and the cut-off velocity, $v_{\rm cut}$, chosen to enclose the appropriate gas mass. These cut-off velocities are then plotted in the {\bf bottom panel} as a function of host halo mass and compared with $v_{\rm c}$. The dotted line shows the theoretically expected scaling with $M_{\rm halo}$.} \label{ansatz}
\end{figure}

The two ideas can be combined. To do this, we consider a momentum-conserving phase, leading to some distribution of outward velocities $\dif M/\dif v_z$, of which only the high-velocity tail, $v>v_{\rm cut}$ is able to escape. This implies:
\eq
M_{out} = \frac{\rm p_{\rm SN}}{\overline{v_z}}\int_{v_{\rm cut}}^\infty\frac{\dif M}{\dif v_z}\dif v_z
\label{general}
\qe
where $p_{\rm SN}$ is the initial momentum provided by the supernovae in question. A more careful derivation considering all directions, rather than our simplistic consideration of outflow directly out of the plane,  would be of similar form but contain structural parameters. 

Further illustration of this approach can be provided by adopting a nominal form for the velocity distribution, $\dif M/\dif v_z$, which appears in  (\ref{general}). We consider the simple form:
\eq
\frac{\dif M}{\dif v_z} = \frac{\frac{1}{2}M_wv_w}{\bar{v}^2}e^{-v_z/\bar{v}}~,\label{dMdv}
\qe 
where $\frac{1}{2}M_wv_w$ has been substituted for the initial momentum provided by the supernovae.

This ansatz for the converged velocity distribution can be compared with measurements from the highest resolution simulations that are available from our range. This comparison is shown in Fig. \ref{ansatz}, which plots the velocity distribution of the ejected gas from three systems of different mass, simulated with the highest number of particles with which all three could be completed. 

The $v_z$ that are measured from the simulations are the eventual exit velocities. Because (\ref{dMdv}) is an attempt to characterise the distribution of velocities at an intermediate stage, these values are adjusted by the potential gain, $\Delta\Phi_i$, at the point of measurement from the pre-kicked position: $v_{\rm adj}=\sqrt{v_z^2+2\Delta\Phi_i}$. This is somewhat crude but is more faithful to the concept behind eqn. \ref{dMdv} and is adequate for the main purpose of this figure, which is to illustrate its application.

The distribution of outflow gas as a function of these adjusted velocities shows that only the high velocity tail makes it out of the disk, and the length of that tail will depend on the potential barrier in question. We model this as a smooth distribution cut-off at at some value, $v_{\rm cut}$, which is determined by the depth of potential well from which the gas has escaped (and thus expected to be close or equal to the characteristic velocity, $v_{\rm c}$). The cut-off is not abrupt in reality, but the approximation of an abrupt cut off may be an effective modelling strategy. 

As shown in Fig. \ref{ansatz}, $v_{\rm cut}$ does indeed scale convincingly with halo mass and $v_{\rm c}$. So the gravitational potential barrier presented to the supernova wind, known to be important since the work of \scite{Matthews71}, can still be included in a simple estimate of the ejected mass, but in a context which is more faithful to the physics of the process.

\subsection{Reconciliation with \S\ref{Introduction}}

\begin{figure}
\includegraphics[trim =  101mm 170mm 15mm 35mm, clip, width=\columnwidth]{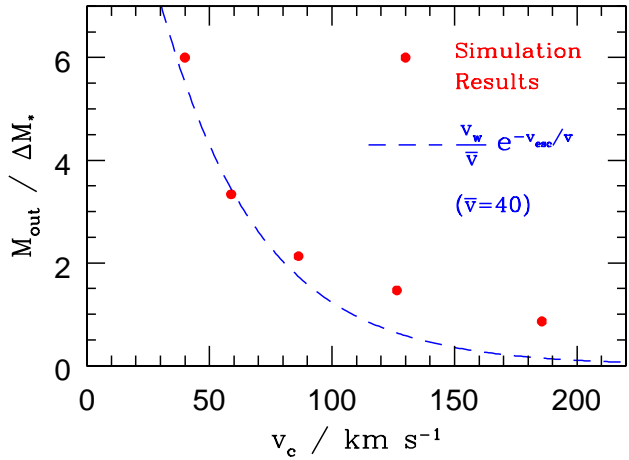}
\caption{A comparison between the simulation results, at the highest achievable resolution, and eqn. \ref{guess}, which attempts to extrapolate beyond the resolution limits. This choice of $\bar{v}$ is based on the velocity distribution which best fits the outflow from the $10^{10}\Msun$ halo, so this can be interpreted as a prediction of the results that {\em would} be found if all the systems were simulated with the particle mass that is possible for this lowest mass galaxy.}\label{check}
\end{figure}

As illustrated in Fig. \ref{ansatz}, eqn. \ref{general} can yield an expression for the outflow in terms of two parameters, both of which have direct physical significance:
\begin{itemize}
\item{The distribution of speeds for supernova shock fronts at the end of their momentum-conserving phase.}
\item{A cut-off velocity, $v_{\rm cut}$, which approximately selects from this distribution the gas mass which will escape the high-density region of a galaxy  (expected to be proportional to $v_{\rm c}$, or correlate with total mass as $M^{2/3}$)}
\end{itemize}
Using the example velocity distribution in (\ref{dMdv}), this gives an alternative equation for the mass outflow:
\ar
\frac{M_{\rm out}}{M_{\rm w}} &\approx& \frac{\frac{1}{2}v_w}{\bar{v}^2}\int_{v_{\rm cut}}^{\infty} e^{-v_z/\bar{v}}~\dif v_z
\nonumber \\
\frac{M_{\rm out}}{M_\star}&\approx& \frac{v_w}{\bar{v}} e^{-v_{\rm cut}/\bar{v}} \label{guess}
\ra
This is compared in Fig. \ref{check} to the results of the highest available resolution runs of the five galaxies (the same points as Fig. \ref{beta}) using values of $\bar{v}=40$ and $v_{\rm cut} = v_{\rm c}$.

The chosen $\bar{v}$ is the value that produced good agreement with the $10^{10}\Msun$ halo, which could be run with the highest resolution (lowest particle mass). Thus, the line in Fig. \ref{check} can be viewed as an attempt to extrapolate from this to predict the outflow that would be found if we could resolve {\em all} the systems to this degree.

For this reason, it is not {\em agreement} with the simulation results that is sought from Fig. \ref{check}. The goal is to supersede them, and to promote a more digestible explanation for observations such as the dependence of feedback efficiency on gravitational potential in {\em real} galaxies.  With this in mind, we return to the issue presented at the start of the paper to see how our analysis of the simulated galaxies might apply in the context of real ones. To begin with, we identify again the two natural limits in the theory. 

\begin{figure}
\includegraphics[trim =  8mm 58mm 74mm 79mm, clip, width=\columnwidth]{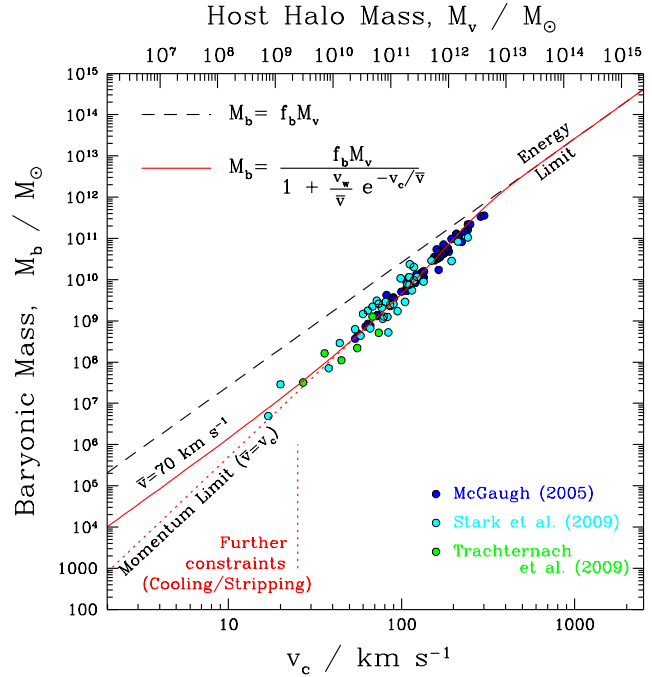}
\caption{Application of the revised theory (\ref{general}) to relevant data from Fig. \ref{McGaugh}. The solid line shows the equation given, using a mean wind velocity of $\bar{v}=70\kms$ and a mean specific initial wind momentum per mass of stars formed of $v_{\rm w}=1400\kms$. The dotted line shows the limiting maximum outflow for this particular wind velocity distribution and value of $v_{\rm w}$.}\label{exp}
\end{figure}

The simplest of these is at high mass, where the escape speed of the system is simply too high and supernovae do not generate the required wind speeds. This is the same physical limit in both theories (eqn. \ref{Larson} and eqn. \ref{general}).

At the very lowest mass, the two models differ. In eqn. \ref{Larson}, the outflow velocity is implicitly set to be zero at large radii and the only limit on the outflow is from energy conservation. But in order to escape at all, the wind must have had at least some outward momentum, of order $M_{\rm out}v_c$, and this momentum must have come from somewhere, implying $M_{\rm out}\ltsim p_{\rm SN}/v_{\rm c}$. The model encapsulated by eqn. \ref{general} does not ignore this, and there is indeed an equivalent maximum outflow for a given momentum yield, $p_{\rm SN}$, from the supernovae. For the velocity distribution of eqn. \ref{dMdv}, this  occurs at $\bar{v}= v_{\rm c}$.

At intermediate masses, there will be a gradual transition between these two limiting physical cases. This is illustrated in Fig. \ref{exp}, which completes the paper by applying the alternative theoretical approach alongside the relevant observational data from Fig. \ref{McGaugh}, which originally motivated the investigation. As indicated in the labelling, the baryon content of the lowest mass structures will be severely limited by additional factors, principally natural cooling thresholds and the stripping of hot gas by larger neighbours.  A detailed explanation of the way these effects function in the context of hierarchical formation can be found in \scite{Stringer10}. For this reason, the low mass observational estimates are not included. The high-mass structures, their baryonic contents also understood to be limited by processes other than supernovae, are similarly omitted.

The values of $v_{\rm w}$ and $\bar{v}$ in Fig. \ref{exp} are chosen to illustrate that the theory is compatible with the observations using physically appropriate parameter values. The value for the momentum yield\footnote{This yield would be available from a supernova sweeping up a mean mass of $~100-300\Msun$ before a significant fraction of an initial energy of $\sim 10^{44}$J was lost through cooling (depending on the IMF and structural factors of order unity).} of $p_{\rm SN}=1400M_\star\kms$ is somewhat higher than the nominal value plotted in Fig. \ref{McGaugh}, but the locus is similar. This is because  the models reviewed in \S\ref{Introduction} consider only one, characteristic value of momentum or energy for the wind, and assume that all the wind escapes. The revised theory allows for a {\em distribution} of velocities in the wind; the slow-moving fraction being retained by gravity, and the high-velocity tail escaping with, in general, a significant asymptotic kinetic energy. We consider this to be a more realistic picture.

\section{Summary}\label{Summary}

In this paper, we began by reviewing the 40 year old theory of supernova-driven outflow from galaxies, distinguishing between the premise of a consistent energy yield and that of a consistent momentum yield. If adapted to CDM cosmology, we demonstrated that such theories can be used to derive 1st-order predictions of the expected baryon content of cosmic structures which are in agreement with observational estimates spanning many orders of magnitude in host halo mass (Fig. \ref{McGaugh}).

With the aid of modern computational resources, we then scrutinised the arguments in these theories,  following the extent to which certain premises apply as parameter values (Fig. \ref{ml}) and particle number (Figs. \ref{m10} and \ref{m12}) were varied in the simulation, identifying regimes dominated by momentum conservation and those where energetic limits take over. One traditional premise which was not upheld in these models was the supposition that all the energy conveyed to the gas would be exhausted in overcoming gravity. The simulated outflows left the system with a range of velocities, comprising a fraction of the total energy comparable to that expended against gravity.

The tendency of the distribution of outward velocities towards a smooth form, as we increased the resolution of the simulations, led us to anticipate the converged form of this distribution. This was used as the basis for a  more cohesive theoretical picture of the process, which takes into account both the local, momentum-conserving properties of the wind and the importance of the gravitational potential barrier which the winds must overcome. 

An example velocity distribution was then presented (Fig. \ref{ansatz}) to demonstrate how such a modelling approach could work, and used to extrapolate predictions for the outflow mass which were compared with the available simulation results (Fig. \ref{check}). The model was also compared with the observational estimates (Fig. \ref{exp}), demonstrating the potential for the emergent global properties of these systems to be accounted for directly in terms of their internal structure and the energy sources which drive them.

\section*{Acknowledgments}

This work was supported by an STFC rolling grant to the Institute for Computational Cosmology. MJS also acknowledges support from the Kavli Institute of Cosmology in Cambridge, and from ESC grant number 267399. CSF acknowledges a Royal Society Wolfson Research Merit Award and an ERC Advanced Investigator grant.  The {\sc gadget-3} code was adapted for this study with the help and permission of  Craig Booth, Rob Crain, Claudio Dalla Vecchia  \&  Joop Schaye and images were generated with John Helly's {\sc gadgetviewer} software. The authors are grateful to Andrew Benson, Francoise Coombes, Stacey McGaugh, Greg Novak, Thanu Padmanabhan \& Phillip Podsiadlowski for all their helpful comments. We also thank the referee for a report which led to considerable additions and improvements to the manuscript.

\appendix

\section{Softening Length}\label{Softening}
In section \ref{Convergence} there is a detailed comparison between simulation results generated using different numbers of particles. In order to make sure that this change in number is solely responsible for any difference in outcome, the gravitational softening length is crudely set to the same value for all calculations. 

The choice, 10pc, was made as it is appropriate for the higher resolution runs and these are results that are actually under the most scrutiny. However, this choice is inappropriate for the lower resolution runs. It is important to verify that the sensitivity of the results to particle mass that was exposed in Figs. \ref{m10} and \ref{m12} might not have been alleviated by more thoughtful choices.

\begin{figure}
\includegraphics[trim = 90mm 100mm 14mm 32mm, clip, width=\columnwidth]{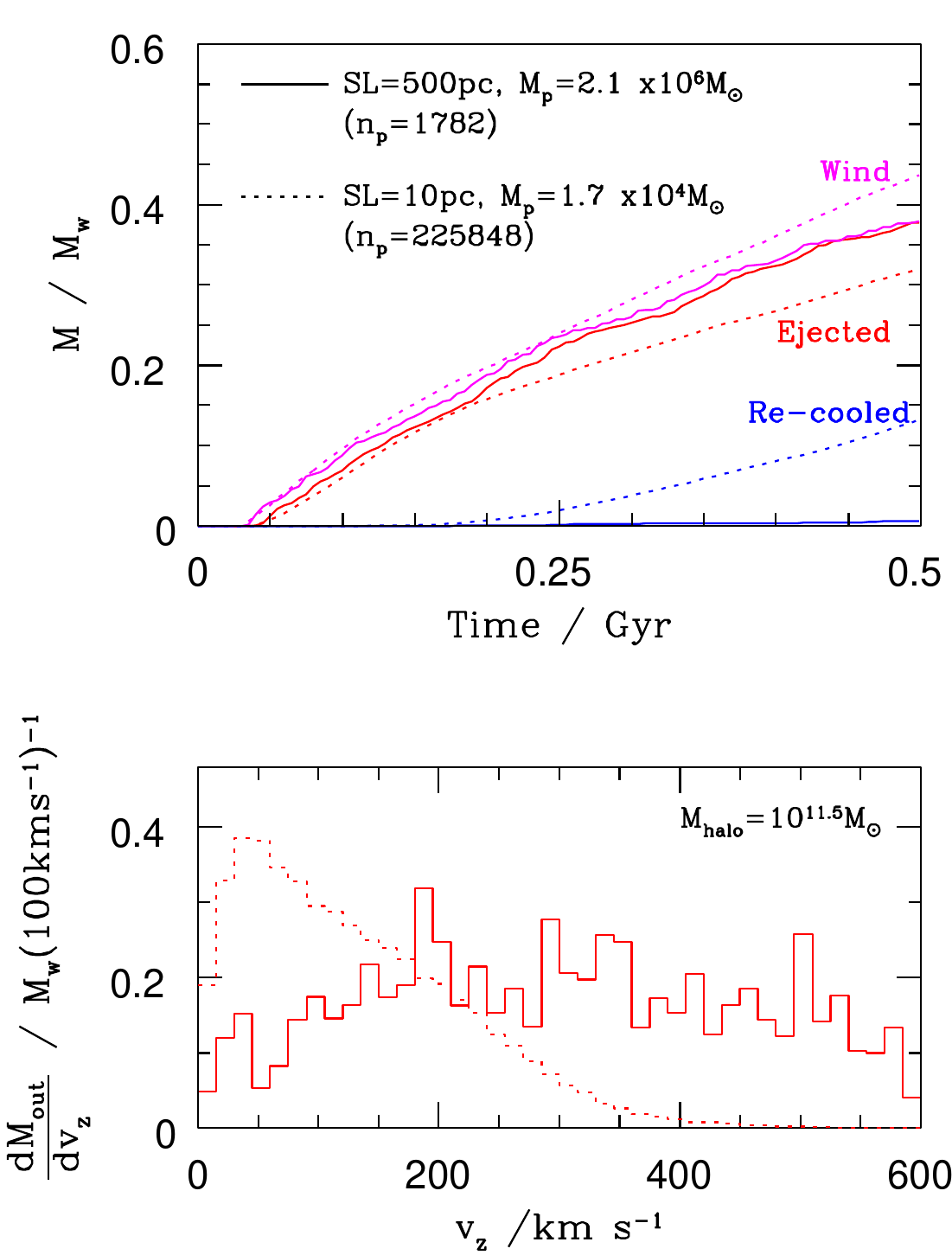}
\caption{Testing the consequences of choosing a more appropriate gravitational softening length for low-resolution simulations. This figure shows the time evolution (upper panel) and exit velocity distribution (lower panel) for  a large disk galaxy at two contrasting resolutions, but with more appropriate choices of softening length than in Figs. \ref{m10} \& \ref{m12}. Whilst the low-resolutions results (solid lines) were indeed sensitive to this change, it clearly does not bring them in line with the high resolution calculations (dotted lines). In the upper panel, the colouring is as for Figs. \ref{m10} \& \ref{m12}: Red for ejected gas, magenta for the initial ``wind'', and blue for the ejected gas that has re-cooled back into the disk.}\label{softening}
\end{figure}

Some basic tests indicated that the results were indeed affected by changing the softening length, but confirmed that no choice could produce identical simulation results when run with very different particle mass. An example of one such test is shown in Fig. \ref{softening}.

\clearpage

\section{Time steps}\label{Timesteps} 

The numerical methods used to illustrate this paper perform integration for each particle using time steps that are determined by its local environment. As pointed out by \scite{Saitoh09}, this will lead to very inaccurate integration in the case of strong shocks, where one locally-determined time step can be very long compared with the that of neighbouring regions.

For the application to which the simulations have been put in this work, we are principally concerned with the question of whether this problem is responsible for the large change in results between low and high particle mass that is shown in Figs. \ref{m10} and \ref{m12}. This possibility was ruled out by running the calculations again for the same initial conditions, but with the code modified such that the timestep for all particles is very short (about 5,000 years). 

\begin{figure}
\includegraphics[trim = 90mm 100mm 14mm 32mm, clip, width=\columnwidth]{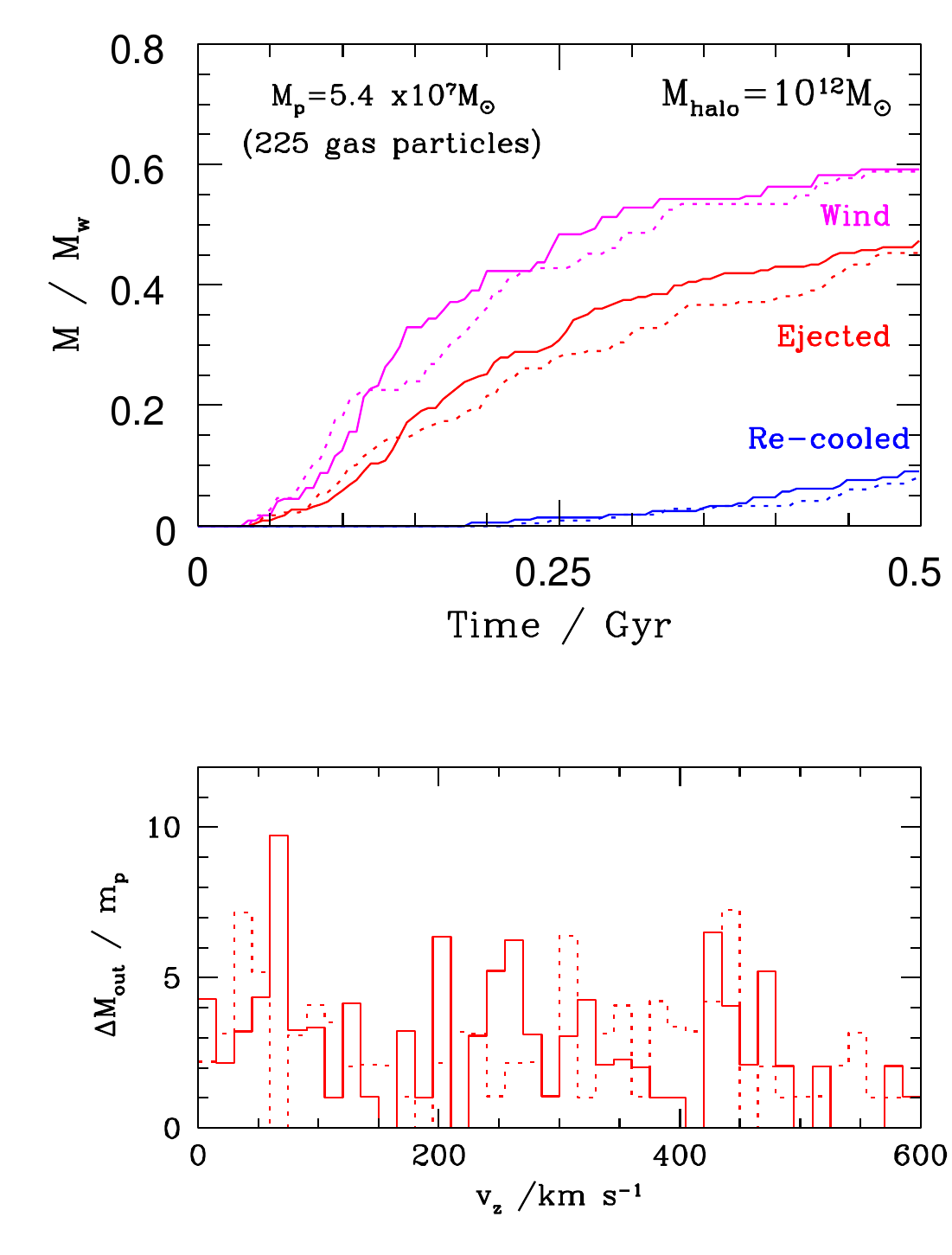}
\caption{Testing the importance of timestep length in the low-resolution simulations. The dotted lines are the results of one of the resolution runs from Fig. \ref{m12}, which used the standard calculation method where timesteps in the calculation are decided dynamically, based on the environment. The solid lines show the evolution of identical initial conditions where the calculation was changed only by enforcing the use of extremely short time steps throughout  ($\Delta t \approx5600$ years). The key is as given, for definitions of the three components, see \S\ref{Simulation}.}\label{tstep}
\end{figure}

The results of one such modified calculation is compared with the results of the standard scheme in Fig. \ref{tstep}. The results are clearly affected by the use of much shorter time steps, but certainly not to the extent that they are changed by the use of smaller particle mass, as shown in Fig.s \ref{m10} and \ref{m12}. Similar tests on other values of halo mass and particle mass produced similar outcomes, supporting the conclusion that rectifying the issues with time step allocation (as achieved by \pcite{Durier11}) would not completely remove the dependence of results on choice of particle mass.

\begin{thebibliography}{widestentry}

\bibitem[Bower et al.<submitted>]{Bower11}
Bower,~R.~,G., Benson,~A.~J., Crain,~R.~A. MNRAS 2011 (accepted)

\bibitem[Chabrier<2003>]{Chabrier03}
Chabrier, G. 2003, PASP, 115, 763

\bibitem[Cole et al.<1994>]{Cole94} Cole, S., Aragon-Salamanca, A., Frenk, C.~S., Navarro, J.~F., 
\& Zepf, S.~E. 1994, MNRAS, 271, 781

\bibitem[Cox<1972>]{Cox72}
Cox, D. P., 1972, ApJ, 178, 159

\bibitem[Crain et al.<2009>]{Crain09} 
Crain, R.~A., Theuns, T., Dalla Vecchia, C., et al.\ 2009, MNRAS, 399, 1773 

\bibitem[Crain et al.<2010>]{Crain10} 
Crain, R.~A., McCarthy, I.~G., Frenk, C.~S., Theuns, T., \& Schaye, J.\ 2010, MNRAS, 407, 1403 

\bibitem[Dalla Vecchia \& Schaye<2008>]{DallaVecchia08} 
Dalla Vecchia, C., \& Schaye, J. 2008, MNRAS, 387, 1431 

\bibitem[De Lucia et al.<2011>]{DeLucia11} De Lucia, G., 
Fontanot, F., Wilman, D., \& Monaco, P.\ 2011, MNRAS, 414, 1439  

\bibitem[Dekel \& Silk<1986>]{Dekel86} 
Dekel, A. \& Silk, J., 1986, ApJ, 303, 39

\bibitem[Durier \& Dalla Vecchia<2012>]{Durier12} 
Durier, F., \& Dalla Vecchia, C.\ 2012, MNRAS, 419, 465 

\bibitem[Font et al.<2011>]{Font11} 
Font, A.~S., McCarthy, I.~G., Crain, R.~A., et al.\ 2011, MNRAS, 416, 2802

\bibitem[Giodini et al.<2009>]{Giodini09} Giodini, S., et al. 2009, ApJ, 703, 982

\bibitem[Hernquist<1990>]{Hernquist90} 
Hernquist, L.\ 1990, ApJ, 356, 359 

\bibitem[Iwamoto et al.<1994>]{Iwamoto94} 
Iwamoto, K., Nomoto, K., Hoflich, P., Yamaoka, H., Kumagai, S., Shigeyama,
T., et al.\ 1994, ApJL, 437, L115 

\bibitem[Komatsu<2011>]{WMAP}
Komatsu, E., et al.\ 2011, ApJS, 192, 18 

\bibitem[Larson<1974>]{Larson74} 
Larson, R.~B. 1974, MNRAS, 169, 229 

\bibitem[McGaugh et al.<2005>]{McGaugh05} 
McGaugh, S. S. 2005, ApJ, 632, 859 

\bibitem[McGaugh et al.<2010>]{McGaugh10}
McGaugh, S.~S., Schombert, J.~M., de Blok, W.~J.~G., \& Zagursky, M.~J. 2010, ApJL, 708, L14 

\bibitem[McKee \& Ostriker<1977>]{McKee77}
McKee~C.~F. \& Ostriker~J.~P., 1977, ApJ,  218, 148

\bibitem[Matthews \& Baker<1971>]{Matthews71}
Matthews, W. G. \& Baker, J. C., 1971, ApJ, 170, 241

\bibitem[Minkowski<1967>]{Minkowski67}
Minkowski, R., 1967, in Middlehurst,~B.~M. \& Aller,~L.~H.,  eds., Nebular and Interstellar Matter. Vol. 7,  University of Chicago Press, Chicago, p. 623 

\bibitem[Murray et al.<2005>]{Murray05} 
Murray, N., Quataert, E., \& Thompson, T.~A.\ 2005, ApJ, 618, 569 

\bibitem[Navarro \& White<1993>]{Navarro93} 
Navarro, J.~F., \& White, S.~D.~M. 1993, MNRAS, 265, 271

\bibitem[Nomoto et al.<1993>]{Nomoto93} 
Nomoto, K., Suzuki, T., Shigeyama, T., Kumagai, S., Yamaoka, H., Saio, H., et al.\ 1993, Nature, 364, 507 

\bibitem[Oppenheimer \& Dav\'{e}<2006>]{Oppenheimer06}
Oppenheimer,~B.~D \& Dav\'{e}, R. 2006, MNRAS, 373, 1265

\bibitem[Saitoh \& Makino<2009>]{Saitoh09} 
Saitoh, T. R. \& Makino, J., 2009, ApJ 697, L99

\bibitem[Schaye \& Dalla Vecchia<2008>]{Schaye08} 
Schaye, J., \& Dalla Vecchia, C.\ 2008, MNRAS, 383, 1210 

\bibitem[Schaye et al.<2010>]{Schaye10} Schaye, J., Dalla 
Vecchia, C., Booth, C.~M., et al.\ 2010, MNRAS, 402, 1536 

\bibitem[Shigeyama \& Nomoto<1990>]{Shigeyama90} 
Shigeyama, T., \& Nomoto, K.\ 1990, ApJ, 360, 242 

\bibitem[Springel<2005>]{Springel05}
Springel, V. 2005 MNRAS, 364, 1105

\bibitem[Springel, Di Matteo \& Hernquist<2005>]{Springel05a}
Springel,~V., Di Matteo, T. \& Hernquist,~L., 2005 MNRAS, 361, 776

\bibitem[Stringer, Cole \& Frenk<2010>]{Stringer10} 
Stringer, M.~J., Cole, S., Frenk, C.~S. 2010, MNRAS, 404, 1129

\bibitem[Trachternach et al. <2009>]{Trachternach09} 
Trachternach, C., de Blok, W. J. G., McGaugh, S. S., van der Hulst, J. M.,
Dettmar, R.-J., 2009, A\&A 505, 577

\bibitem[ Walker et al. <2009>]{Walker09} 
Walker, M. G.; Mateo, M., Olszewski, E. W., Pe–arrubia, J., Wyn Evans, N.,
Gilmore, G., 2009 ApJ, 704, 1274 

\bibitem[White \& Frenk<1991>]{White91}
White, S. D. M. \& Frenk, C. S. 1991, ApJ, 379, 52

\bibitem[White \& Rees<1978>]{White78}
White, S. D. M. \& Rees, M. J. 1978, MNRAS, 183, 341

\bibitem[Wiersma et al.<2009>]{Wiersma09} 
Wiersma, R.~P.~C., Schaye, J., \& Smith, B.~D.\ 2009, MNRAS, 393, 99 

\end{thebibliography}
\end{document}